\documentclass[12pt]{article}

\usepackage{amsmath}
\usepackage{amssymb}
\usepackage{amsthm}
\usepackage{mathtools}
\usepackage{float}
\usepackage{geometry}
\usepackage{wrapfig}
\usepackage[font=small,skip=2pt]{caption}
\usepackage{xcolor}
\usepackage[hidelinks]{hyperref}
\usepackage{url}
\usepackage{array}
\usepackage{wrapfig}
\usepackage{etoolbox}
\usepackage{marginnote}
\usepackage{multicol}
\usepackage[skip=0pt plus1pt, indent=20pt]{parskip}
\usepackage{enumitem}
 \setlist[itemize]{noitemsep, topsep=0pt}
\usepackage{setspace}
\newcommand{\myspace}{\vspace{7pt}}
\newcommand{\mysubsubsection}[1]{
\vspace{-7pt}
\subsubsection{#1}
\vspace{-7pt}
}
\newcommand{\mysection}[1]{
\vspace{-12pt}
\section{#1}
\vspace{-8pt}
}
\newcommand{\mysubsection}[1]{
\vspace{-7pt}
\subsection{#1}
\vspace{-7pt}
}
\newcommand{\myparagraph}[1]{
\paragraph{#1} 
\vspace{-7pt}
}

\begin{document}

%%%%%%%%%%   begin document & title page      %%%%%%%%%% 
\newpage
\clearpage
\pagenumbering{arabic} 
\begin{center}

\vspace*{50pt}
\noindent \textbf{{\Large Off-resonance artifact correction for magnetic resonance imaging: a review}}

\vspace*{20pt}

\noindent Melissa W. Haskell\textsuperscript{1,2}, Jon-Fredrik Nielsen\textsuperscript{3}, Douglas C. Noll\textsuperscript{3}

\vspace*{20pt}

\noindent {\footnotesize \textsuperscript{1}Electrical Engineering and Computer Science, University of Michigan, Ann Arbor, MI, United States\\
\vspace{15pt}
\textsuperscript{2}Hyperfine Research, Guilford, CT, United States \\
\vspace{15pt}
\textsuperscript{3}Biomedical Engineering, University of Michigan, Ann Arbor, MI, United States}

\end{center}

\vspace*{50pt}

\providetoggle{annotated}
\settoggle{annotated}{false}

\section*{Abstract}
\addcontentsline{toc}{section}{Abstract}

In magnetic resonance imaging (MRI), inhomogeneity in the main magnetic field used for imaging, referred to as off-resonance, can lead to image artifacts ranging from mild to severe depending on the application. Off-resonance artifacts, such as signal loss, geometric distortions, and blurring, can compromise the clinical and scientific utility of MR images. In this review, we describe sources of off-resonance in MRI, how off-resonance affects images, and strategies to prevent and correct for off-resonance. Given recent advances and the great potential of low field and/or portable MRI, we also highlight the advantages and challenges of imaging at low field with respect to off-resonance.

%%%%%%%%%%   sections      %%%%%%%%%% 
\newpage

\mysection{Introduction}

Magnetic resonance imaging (MRI) is an important clinical tool for diagnosis and intervention, as well as the leading modality for noninvasively imaging the \emph{in vivo} human brain. At the core of the MR imaging experiment is a polarizing magnetic field, referred to as the $B_0$ field or the main magnetic field, that aligns the magnetic spins within an object. These spins will precess at a resonant frequency linearly proportional to the external $B_0$ field, and in the majority of MRI experiments it is ideal to have as homogeneous a $B_0$ field as possible so that all spins precess at the same frequency. In practice, many factors can lead to $B_0$ inhomogeneity, thus leading to a non-uniform precession frequency of the spins, and this is called \emph{off-resonance}. Off-resonance can lead to artifacts in MR images, including signal loss, geometric distortions, and blurring. 

Off-resonance artifacts have been reported since the early days of clinical MRI, along with the problems they present in interpreting MR images~\cite{Ludeke1985SusceptibilityImaging,ODonnell1985NMRNonlinearities,Czervionke1988MagneticImaging.}. Geometric distortions that warp the shape of anatomy create challenges in many clinical applications, including stereotactic localization~\cite{Michiels1994OnNeurosurgery}, MRI-guided biopsy~\cite{Ladd1996BiopsyArtifacts}, and MRI-guided radiation therapy~\cite{Walker2014MRIPlanning,Weygand2016SpatialDistortion}. Metal implants can lead to severe geometric distortions, and also can create a total loss of signal around the metal object~\cite{Hargreaves2011Metal-inducedMRI}. In MRI based neuroscience research, uncorrected off-resonance artifacts can lead to a loss in quality of data, most notably in the field of functional MRI (fMRI)~\cite{Jezzard1999SourcesData, Jezzard2012CorrectionData}. 

Because of these problems, many approaches exist in MRI workflows to mitigate off-resonance artifacts. Magnetic field shimming, where dedicated shim coils apply additional fields to offset and cancel out main field inhomogeneity, is automated and standard on most vendors. From a sequence perspective, one of the simplest solutions to avoiding signal loss from off-resonance is to use spin echo-based sequences~\cite{Hahn1950SpinEchoes,Hennig1986}, which refocus signal at the echo time (TE). The expanded use of parallel imaging~\cite{Larkman2004,Feinberg2013Ultra-fastImaging} has also allowed for smaller voxels in standard imaging protocols. Smaller voxels reduce signal loss because the magnetic fields within the voxels have less variation in frequency, hence less signal cancellation. There is also less $B_0$ distortion with parallel imaging, because data can be acquired in shorter readouts with higher effective bandwidths in k-space.

However, not all $B_0$ artifacts can be avoided with these approaches. 
The magnetic field off-resonance map, called a $B_0$ \emph{fieldmap} or \emph{fieldmap} for short, inside the human body often has spatial variations that are much higher order than the 2$^\text{nd}$ order spherical harmonic shim corrections available on some systems, leaving the potential for residual off-resonance in hard to shim anatomy.
Additionally, certain sequences require gradient echo imaging with longer readouts to generate the desired image contrast or acquisition speed, and are therefore more prone to off-resonance artifacts (e.g., echo planar imaging (EPI)~\cite{Stehling1991Echo-PlanarSecond}). Longer TEs are important for sequences such as blood oxygen level dependent (BOLD) contrast fMRI~\cite{Kwong1992DynamicStimulation}, to allow for dephasing of the spins to present as $T^*_2$ contrast. Susceptibility weighted imaging (SWI) and quantitative susceptibility mapping (QSM) also require longer TEs to generate magnetic susceptibility contrast, and these sequences are useful tools to investigate pathologies such as intracranial hemorrhage, traumatic brain injury, stroke, neoplasm, and multiple sclerosis~\cite{Liu2015Susceptibility-weightedBrain,Duyn2017ContributionsTissue}. 
Further, even when the desired contrast does allow for the use of spin-echo imaging and most $B_0$ induced signal loss can be avoided, geometric distortions can still be present due to unwanted gradients that add to the imaging gradients,  resulting in k-space encoding errors.

It is also important to consider off-resonance artifacts and potential correction strategies when working with MRI scanners that have more inhomogeneous $B_0$ fields than high-field superconducting magnets. Recently, there have been many advances in portable MRI at lower fields to perform \emph{in vivo} human imaging~\cite{Nakagomi2019DevelopmentMagnet,Liu2021AScanner,Cooley2021ABrain,OReilly2021InArray,Sheth2021AssessmentPatients}. These new technologies are important for increasing the accessibility of MRI~\cite{Geethanath2019AccessibleReview}, but they generally have less uniform $B_0$ fields than superconducting magnets and off-resonance correction approaches are needed. Low field MRI scanners built with permanent magnets can also experience large field drifts due to temperature induced field changes, so dynamic tracking of the $B_0$ field strength is sometimes needed during imaging. 

This review is structured by first describing the sources of off-resonance in MR imaging in Section~\ref{sec:sources} and their effects on imaging in Section~\ref{sec:imaging}. Section~\ref{sec:measuringB0} details ways to estimate a $B_0$ fieldmap using pulse sequence, hardware, and optimization based approaches. Section~\ref{sec:corrections} discusses strategies to prevent and correct for $B_0$ artifacts, and Section~\ref{sec:conclusion} concludes the paper. All code for the simulations shown here is available online at \url{https://github.com/fmrilab/B0-review-2022}, including example code to perform model-based image reconstruction~\cite{Fessler2020OptimizationAlgorithms} that incorporates a $B_0$ fieldmap~\cite{Sutton2003} using the Michigan Image Reconstruction Toolbox (MIRT)~\cite{Fessler}.

\section{Sources of magnetic field off-resonance} \label{sec:sources}

%%%%%%%%%%%%%%%%%%%%%%%%%%%%%%%%%%%%%%%%%%%%
%%%%%%%%   B0 inhomogeneity    %%%%%%%%%%%%%
\mysubsection{Main field inhomogeneity}
\myspace % needed when a "my" command comes right after another one
\mysubsubsection{Modern superconducting magnet homogeneity}
The $B_0$ polarizing fields of modern high-field superconducting magnets, most often at field strengths of 1.5T or 3T, are generally very uniform, and after shimming with superconducting shim coils and passive shim elements are on the order of 1ppm uniformity over a large imaging volume. Because of this, artifacts from susceptibility, chemical shift, and metal implants (discussed in later sections) are the dominant sources of $B_0$ artifacts in high-field superconducting MRI scanners, and off-resonance artifacts from the magnet itself are minimal.

\mysubsubsection{New magnet designs for lower cost and portability}
As more light-weight, low-cost, and portable MRI designs are being developed, permanent magnets and resistive electromagnets are being chosen to create the polarizing fields. Both permanent and resistive magnet designs are much less expensive to produce than superconducting magnets, and, importantly, are much less expensive and easier to operate and maintain. However, these magnets are generally less spatially homogeneous and may require $B_0$ field correction even for routine imaging. Permanent magnets have the advantage that they require no power or cooling mechanism to operate, but they can be very sensitive to temperature induced field drifts, which often need to be monitored during scanning. Resistive magnets are not as temperature sensitive, but power consumption and cooling will need to be taken into consideration and will potentially limit the field strength and homogeneity. 

There are examples both in industry and academia of newer permanent magnetic designs. Currently available commercially scanners include a 64 mT portable brain scanner (\url{https://www.hyperfine.io}), a 1T scanner for the neonatal intensive care unit (\url{https://www.aspectimaging.com}), and a low-field system for dedicated prostate imaging and biopsy guidance (\url{https://promaxo.com}). 
Academic groups have created permanent magnet designs using cylindrical Halbach geometries~\cite{Cooley2014,Cooley2018,OReilly2019Three-dimensionalMagnet}, dipole magnet geometries~\cite{Liu2021AScanner,Nakagomi2019DevelopmentMagnet}, and single sided designs~\cite{McDaniel2019TheImaging}. Unlike superconducting magnets, these scanners have field homogeneities in the range of tens of ppm (for smaller FOV extremity imaging) to hundreds or thousands of ppms for head imaging, but have been shown to successfully image human subjects~\cite{Cooley2021ABrain,OReilly2021InArray,Liu2021AScanner,Nakagomi2019DevelopmentMagnet}. 

Resistive magnets at low field have also been explored. Obungoloch et al. designed and constructed a prepolarized MRI scanner that acquired phantom images at 2.66mT with a polarizing field of 27mT, and a field homogeneity of 5000 ppm, similar to the level of inhomogeneity in the permanent magnet Halbach designs~\cite{Obungoloch2018DesignHydrocephalus}.

%%%%%%%%%%%%%%%%%%%%%%%%%%%%%%%%%%%%%%%%%%%%
%%%%%%%%  Susceptibility  %%%%%%%%%%%%%
\mysubsection{Magnetic susceptibility}\label{subsec:sus}
One of the largest sources of $B_0$ inhomogeneity during routine \emph{in vivo} MRI is magnetic susceptibility. Magnetic susceptibility (sometimes called volume magnetic susceptibility)~\cite{Schenck1996TheKinds}, denoted as $\chi$, is the property of a material to either strengthen ($\chi > 0$) or weaken ($\chi < 0$) an applied magnetic field, based on its internal magnetic polarization, $M$, as follows:
\begin{equation}
    \chi = M / B_0
\label{eqn:chi}
\end{equation}
Magnetic susceptibility is unitless since it is the ratio of two fields, but it is often expressed in parts per million (or ppm) for convenience in the context of MRI. For example, molecular oxygen, O$_2$, has a magnetic susceptibility of roughly +1ppm, which means that when in the presence of a 1.5T, or 64 MHz, external magnetic field, its internal polarization will be roughly 1.5$\mu$T, or 64 Hz.

If an object has non-uniform magnetic susceptibility, it will lead to a non-uniform magnetic field, even within the most uniform of applied fields generated by today's modern superconducting magnets. In human imaging, most biological tissues are weakly diamagnetic ($\chi < 0$) and oppose the applied field, but oxygen, present in the lungs, sinuses, and other air cavities, is weakly paramagnetic ($\chi > 0$) and strengthens the applied field. This leads to spatially varying $B_0$ fields within the body, such as the one shown in Fig.~\ref{fig:b0_brain}.

Looking at Eq.~(\ref{eqn:chi}), one can see that as the applied magnetic field $B_0$ increases, the internal polarization, $M$, will also increase. This means that at higher field strengths the amount of off-resonance will increase, and therefore the susceptibility artifacts will be more severe~\cite{Farahani1990EffectImaging}.  
For a similar \emph{in vivo} fieldmap at 3T to the 7T fieldmap shown in Fig.~\ref{fig:b0_brain}, as well as the 3T vs. 7T \emph{in vivo} images with corresponding amounts of susceptibility artifacts, we point the reader to Fig. 5 of Stockmann \& Wald, 2018~\cite{Stockmann2018InT}.
Due to the scaling with field strength, it is sometimes convenient to describe field susceptibility in terms of ppm, where the typical fieldmap often less than 1 ppm for most of the brain, but can grow to about 2 ppm at air/brain interfaces.

\begin{figure}[H]
\begin{center}
\begin{tabular}{ | m{4cm} | m{4cm}| m{4cm} | m{4cm} | } 
  \hline
  \textbf{Magnetic Property} & \textbf{Direction of $M$ Relative to External Field} & \textbf{Relative Magnetic Susceptibility ($\chi$, in ppm)} & \textbf{Example Materials} \\ 
  \hline
  Diamagnetism & Opposite & -10 & Water,  most biological tissues \\ 
  \hline
  Paramagnetism & Same & +1 & Molecular oxygen, O$_2$ \\ 
  \hline
  Superparamagnetism & Same & +5000 & SPIO contrast agents \\ 
  \hline
  Ferromagnetism & Same & $>$10,000 & Iron, steel \\ 
  \hline
\end{tabular}
\end{center}
\caption{\textbf{Magnetic Susceptibility.} \footnotesize{Adapted from \url{https://mriquestions.com/what-is-susceptibility.html} \cite{Elster2022Https://mriquestions.com/what-is-susceptibility.html}}} 
\label{fig:mag_sus}
\end{figure}

\begin{figure}[h]
\centering
\includegraphics[width=\textwidth]{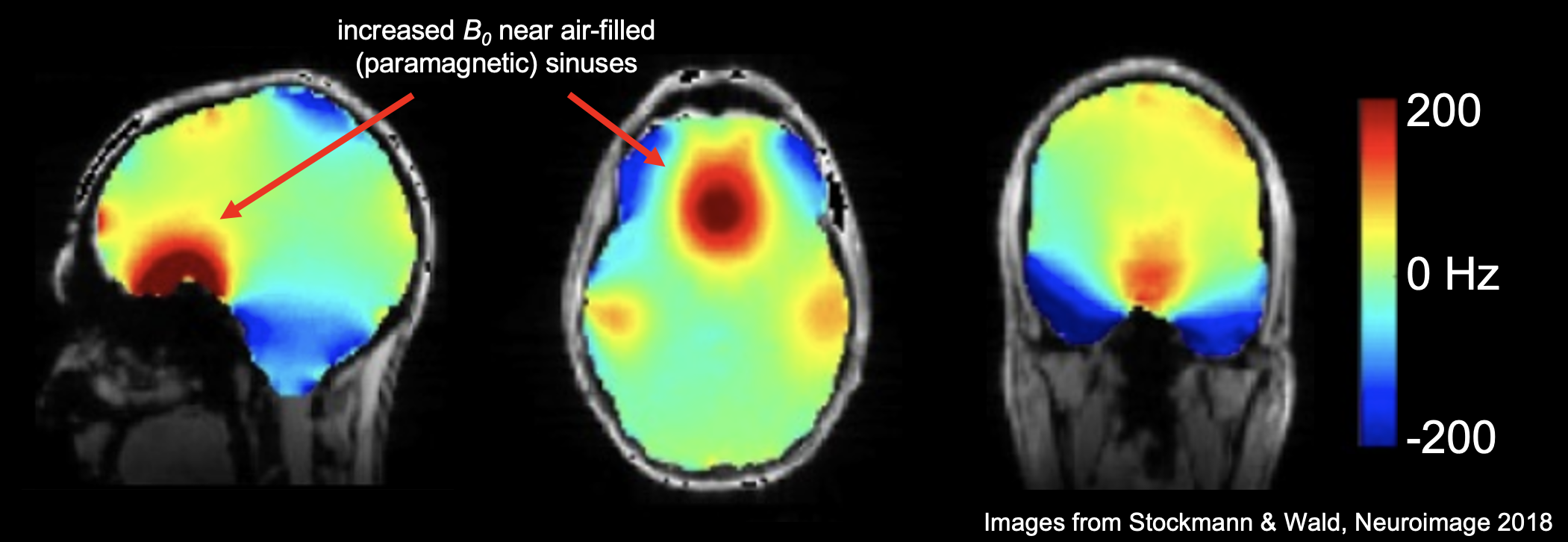}
\caption{\textbf{$\boldsymbol{B_0}$ magnetic field in the brain at 7T with 2$^\text{nd}$ order shimming.} \footnotesize{
This fieldmap shows the spatially varying patterns of the $B_0$ field in the human brain (superimposed over an anatomical image). Even after 2$^\text{nd}$ order shimming has been applied to flatten the field, higher order field inhomogeneities persist. Arrows show that near the air-filled, and therefore paramagnetic, sinuses, the $B_0$ field is higher than areas in the center of the mostly diamagnetic brain tissue.  \emph{(Images reused from Stockmann \& Wald, 2018~\cite{Stockmann2018InT} with permission.)}}} 
\label{fig:b0_brain}
\end{figure}

\begin{wrapfigure}{R}{.5\textwidth}
\centering
\includegraphics[width=.5\textwidth]{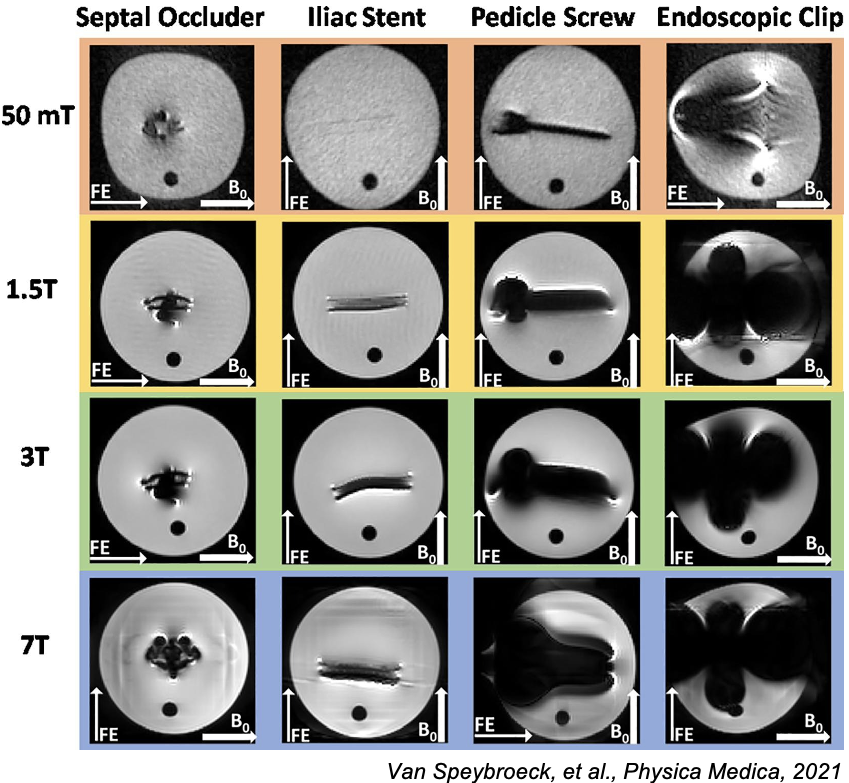}
\caption{\textbf{Metal artifact from 50mT to 7T.} \footnotesize{ Here we show Figure 7 from Van Speybroek, et al., 2021~\cite{VanSpeybroeck2021CharacterizationSystems} where the ``worst-case" scenario images for four types of metal implants are demonstrated. \emph{(Images reused via open access license.)} }} 
\label{fig:metal_50mT_to_7T}
\end{wrapfigure}

%%%%%%%%%%%%%%%%%%%%%%%%%%%%%%%%%%%%%%%%%%%%%%%%%
%%%%%%%%  chemical shift  %%%%%%%%%%%%%
\mysubsection{Chemical shift}
Another source of off-resonance is the chemical shift in resonant frequency of different tissues in the body. The most relevant source of chemical shift is from fat, which has a shift in frequency of -3.5ppm relative to water~\cite{Hood1999ChemicalRevisited}. At 1.5T (64 MHz), this corresponds to a shift in frequency of 64MHz $\times$ -3.5ppm $=$ -224 Hz, and at 3T a shift of -448 Hz. The earliest accounts of chemical shift image artifacts describe enhanced or diminished borders of organs within the body that have tissue boundaries containing fat~\cite{Babcock1985EdgeEffect, Dwyer1985FrequencyImaging}. This is due to the fat signal shifting within the image because it is off-resonant compared to water (Eq.~(\ref{eqn:dx})). The directions and angles of the slice selection and frequency encoding axes can also affect chemical shift artifacts, leading to them sometimes being present and sometimes not, which can confound diagnosis and affect treatment~\cite{Smith1991ChemicalInterface.}. 
Hood et al.~\cite{Hood1999ChemicalRevisited} provide a good review of how chemical shift artifacts impact clinical interpretation of images. The authors also describe what are referred to as chemical shift artifacts of the 2$^\text{nd}$ kind, which are not shifts in image space of the voxel, but rather a cancellation of the out-of-phase water and fat components in the voxel that leads to a loss of total signal (see Hood et al. Fig.~10). These out-of-phase images at a given TE can be compared to images with different TEs where fat is visible, allowing the clear identification of fat.

%%%%%%%%%%%%%%%%%%%%%%%%%%%%%%%%%%%%%%%%%%%%%%%%%
%%%%%%%%  metal implants %%%%%%%%%%%%%
\mysubsection{Metal implants}

While magnetic susceptibility differences drive many sources of $B_0$ inhomogeneity as described in Section \ref{subsec:sus}, the $B_0$ effects of metal implants are different, mainly due to the sheer magnitude of the $B_0$ shifts, leading to effects that often require advanced measures to compensate.  Most metal implants in common use today are labeled as MRI-conditional, which indicates that MRI scanning is acceptable provided the specified conditions for safe use are met~\cite{ASTM2013ATSMEnvironment.,IEC2014IECEnvironment.}.  Allowable or safe use, however, does not mean that the quality of the images is not impacted by susceptibility effects around the implant.  

Hagreaves et al.~\cite{Hargreaves2011Metal-inducedMRI} is a review on sources and artifacts due to metal implants as well as a compendium on the various approaches to mitigate the artifacts.  Jungmann et al.~\cite{Jungmann2017AdvancesMetal} also reviews mitigation methods and provides numerous imaging examples. As with all susceptibility-induced $B_0$ changes, these effects scale with field strength.  Notably, lower magnetic fields result in smaller distortions and lower field strength clinical scanners are often preferred when scanning structures proximal to metal implants, as seen in Fig.~\ref{fig:metal_50mT_to_7T}.  Further, the lower resonant frequency at lower field reduces the safety and imaging impact of RF ($B_1$) inhomogeneity around metal implants.

\section{Impact on Imaging}\label{sec:imaging}

%%%%%%%%%%%%%%%%%%%%%%%%%%%%%%%%%%%%%%%%%%%%%%%%%%%%%%%%%%
%%%%%%%%  ideal B0 and some basic MR physics %%%%%%%%%%%%%
\mysubsection{MRI Signal Equation Incorporating Off-Resonance}
In an MRI experiment, the magnetic spins being imaged precess at resonant frequency $\omega$:
\begin{equation}
    \omega = \gamma B
\end{equation}
where $\gamma$ is a spin's gyromagnetic ratio and $B$ is the magnetic field strength at the spin location. For an $^1$H proton in water for which $\gamma/ 2\pi=42.577$ MHz/T, at a magnetic field strength $B=$ 3T, the spins precess at roughly 128 MHz. Note that throughout this paper we may use ``field" or ``frequency" depending on context, but the reader can convert from a precession frequency in Hz to a magnetic field in T or vice versa by multiplying or dividing by $\gamma/2\pi$.

To create an image in an ideal MRI experiment, first, the object is placed in a spatially uniform polarizing field, $B_0$. The $^1$H spins throughout the entire object will now be precessing at $\omega_0$. Next, spatially varying magnetic fields, $G_x$, $G_y$, and $G_z$, referred to as ``gradient fields" or simply ``gradients", are applied to linearly change the magnetic field as a function of spatial position, which therefore changes the precession frequency of spins at different locations, i.e., $\omega(x,y,z) = \gamma (B_0 + G_x x + G_y y + G_z z) $. By applying different gradient fields over time, spins at different locations can be encoded, the most common method being Fourier space, or $k$-space, encoding. The MRI signal as a function of these time-varying gradients can be described using the signal equation:
\begin{equation}
s(t) = \int m(x,y) e^{-i2\pi [k_x(t)x+k_y(t)y]} dxdy \hspace{10pt} \text{,} \hspace{10pt} k_{\alpha}(t) = \frac{\gamma}{2\pi} \int_{0}^{\tau} G_{\alpha}(\tau) d\tau 
\label{eqn:signal}
\end{equation}
where $s(t)$ is the time-varying MRI signal, $m(x,y)$ is the magnetization at location $(x,y)$, and $k_x(t)$ and $k_y(t)$ are the $k$-space sampling locations. For brevity, the equation above and the rest of the image encoding mathematics will be in 2D, but it can be extended to 3D as well by adding a third $z$ term to the exponent in Eq.~(\ref{eqn:signal}).

By having a spatially uniform polarizing $B_0$ field, the time-varying signal $s(t)$ is only a function of the applied gradients. 
\iftoggle{annotated}{\textcolor{blue}{However when $B_0$ is not spatially uniform, i.e., $B_0(x,y) = B_{0,ideal} + \Delta B_0(x,y)$, \marginnote{\textcolor{blue}{RC}} we must add a term to the signal equation for the \emph{additional} (and generally unwanted) spatially dependent phase, $\phi(x,y)$, accrued over time $t$ due to $\Delta B_0(x,y)$}}{However when $B_0$ is not spatially uniform, i.e., $B_0(x,y) = B_{0,ideal} + \Delta B_0(x,y)$, we must add a term to the signal equation for the \emph{additional} (and generally unwanted) spatially dependent phase, $\phi(x,y)$, accrued over time $t$ due to $\Delta B_0(x,y)$}:
\begin{equation}
    \phi(x,y,t) = \gamma \iftoggle{annotated}{\textcolor{blue}{\Delta B_0(x,y) t = \Delta \omega_0 (x,y) t = 2\pi \Delta f_0(x,y)t}\marginnote{\textcolor{blue}{RC}}}{{\Delta B_0(x,y) t = \Delta \omega_0 (x,y) t = 2\pi \Delta f_0(x,y)t}}  
    \label{eqn:B0phase}
\end{equation}

The phase can now be added to the signal equation using a complex exponential of the product of the fieldmap and the time, $t$:
\begin{equation}
s(t) = \int m(x,y)
\iftoggle{annotated}{\textcolor{blue}{e^{-i\Delta \omega_0(x,y)t}}\marginnote{\textcolor{blue}{RC}}}{e^{-i\Delta \omega_0(x,y)t}} e^{-i2\pi [k_x(t)x+k_y(t)y]} dxdy
\label{eqn:signal_omega}
\end{equation}
This leads to a breakdown of the encoding properties of the MRI signal equation, since now there will be additional phase in the object that is not a result of the $k$-space encoding. Depending on the length of the readout, the type of $k$-space acquisition, and other scan parameters, the excess phase accrued due to off-resonance can lead to various types of artifacts such as signal loss, distortions, and blurring.

\begin{figure}[h]
\centering
\includegraphics[width=\textwidth]{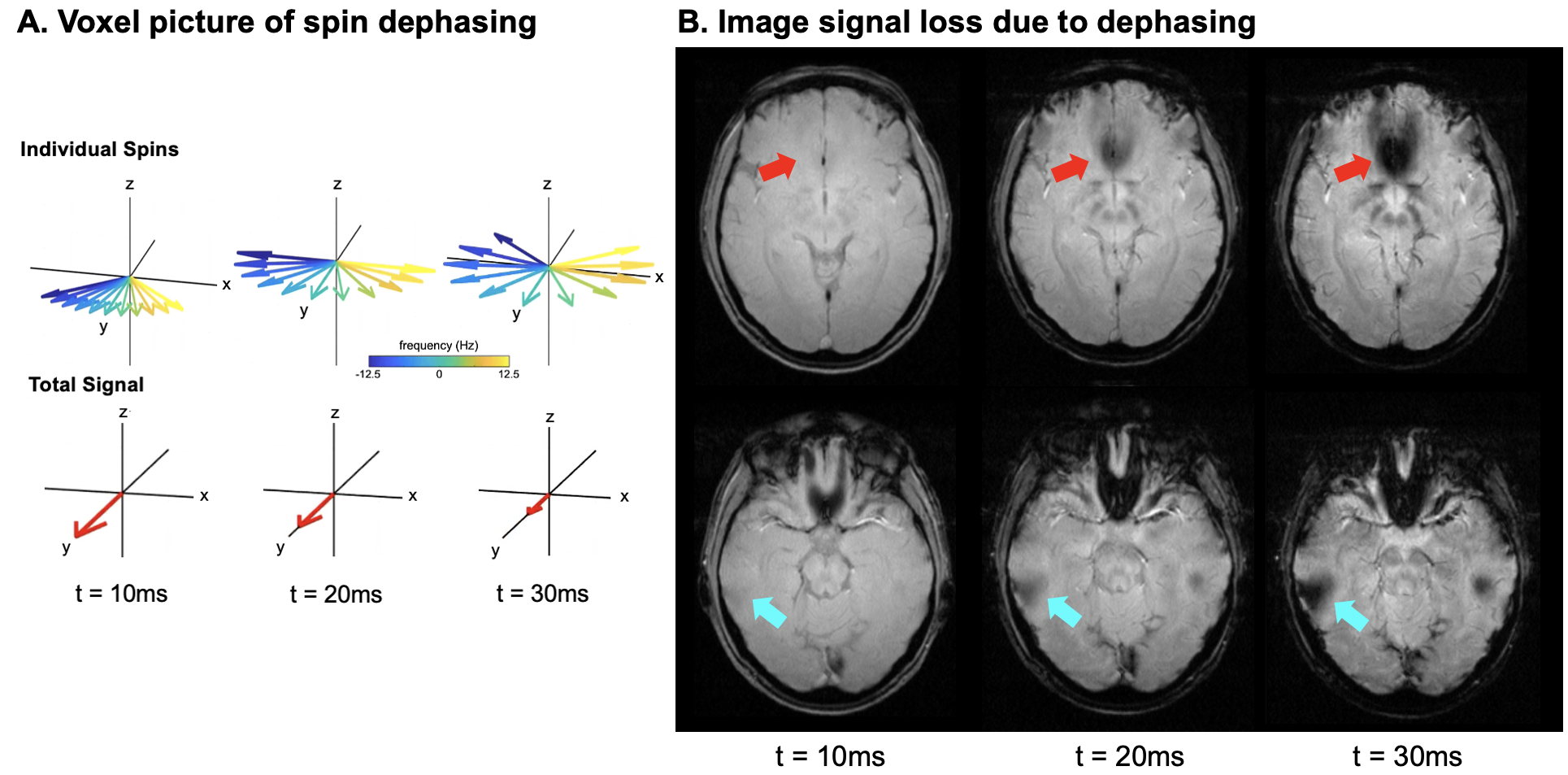}
\caption{\textbf{Signal Loss from Spin Dephasing.} \footnotesize{
A. Voxel view of spin dephasing. Spins with different off-resonance will gain different amounts of phase during signal readout leading them to ``fan out", and eventually completely cancel each other. B. A gradient echo image is shown for two different slices at three different echo times (TE). At the later echo times, signal loss due to dephasing is present as indicated by the arrows.
}} 
\label{fig:dephase}
\end{figure}

%%%%%%%%%%%%%%  signal loss  %%%%%%%%%%%%%%%%%%%%%%%%%%%
\noindent \textbf{Signal Loss}

\noindent When spins in a voxel have different resonant frequencies due to rapid spatial variations in the $B_0$ field, spins will dephase as seen in Fig.~\ref{fig:dephase} (images in Fig.~\ref{fig:dephase} were acquired from a healthy subject after obtaining IRB approved written informed consent). Immediately after RF excitation the spins are all aligned, but during the course of the readout the spins dephase and there is destructive interference in the voxel, referred to as $T^*_2$ decay. The most common way to deal with this signal loss is through the use of spin echoes~\cite{Hahn1950SpinEchoes}. In a spin echo sequence, the spins are flipped into the transverse plane, and then after time TE$/2$, a 180$^{\circ}$ RF pulse flips the spins about one of the transverse axes. Then, after precessing for time TE/2, the spins will refocus at the echo time, TE. Spin echoes are very robust overall, and because of this  spin echo based sequences have been the pulse sequence of choice for many of the academic low field scanners~\cite{Cooley2021ABrain,OReilly2021InArray,Liu2021AScanner,Obungoloch2018DesignHydrocephalus}, as well as the commercially available 64 mT portable brain scanner~\cite{Sheth2021AssessmentPatients}.

%%%%%%%%%%%%%%  encoding artifacts  %%%%%%%%%%%%%%%%%%%%%%%%%%%
\noindent \textbf{Image Encoding Artifacts: Distortions \& Blurring}

\begin{wrapfigure}{r}{.5\textwidth}
\includegraphics[width=.5\textwidth]{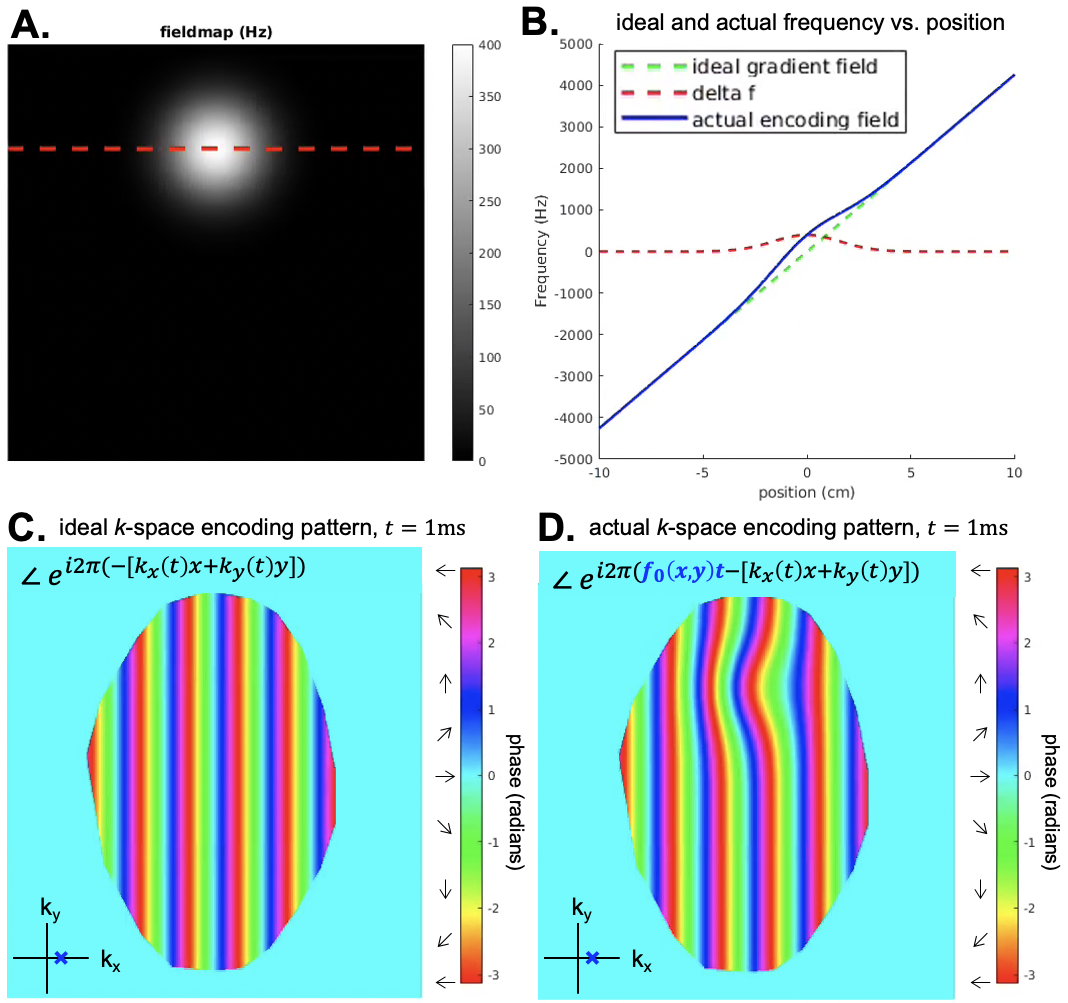}
\caption{\iftoggle{annotated}{\textcolor{blue}{\textbf{Ideal vs. actual encoding phase at a single k-space location.}}}{\textbf{Ideal vs. actual encoding phase at a single k-space location.}} \footnotesize{
\textbf{A}. Simulated off-resonance fieldmap in Hz.
\textbf{B.} 1D plots of the encoding fields as a function of position, where the red dotted line correspond to the dotted red line in A. 
\textbf{C.} The ideal k-space encoding phase for $(k_x,k_y)=(0,0.426 \text{ cm}^{-1})$. 
\iftoggle{annotated}{\textcolor{blue}{\textbf{D.} The actual k-space encoding phase for $(k_x,k_y)=(0,0.426 \text{ cm}^{-1})$. Combined with k-space encoding errors at other k-space sampling locations, this phase  error will lead to off-resonance image artifacts.}}{\textbf{D.} The actual k-space encoding phase for $(k_x,k_y)=(0,0.426 \text{ cm}^{-1})$. Combined with k-space encoding errors at other k-space sampling locations, this phase error will lead to off-resonance image artifacts.}
}} 
\label{fig:encoding}
\iftoggle{annotated}{\marginnote{\textcolor{blue}{RC}}}{}
\end{wrapfigure}

\noindent Spin echoes can mitigate much of the off-resonance based signal loss, but even using spin echo sequences, significant off-resonance artifacts can be present based on image encoding errors, such as distortion~\cite{Wang2005GeometricImaging}, blurring, and pile-up. Looking at Eq.~(\ref{eqn:signal_omega}), the reduction of intravoxel dephasing by using a spin echo leads to an apparent increase in the $m(x,y)$ term, but the spatial encoding in the exponential of that equation will still be affected by $B_0$ effects. When there is a non uniform fieldmap like the one in Fig.~\ref{fig:b0_brain}, there is a breakdown in the encoding since it is not only scanner gradient $k$-space encoding that is contributing to a voxel's phase, but also phase accrued due to off-resonance. We can factor out 2$\pi$ and combine the exponentials into one term as follows:
\begin{equation}
s(t) = \int m(x,y) 
\iftoggle{annotated}{\textcolor{blue}{e^{-i2\pi[\Delta f_0(x,y)t + [k_x(t)x+k_y(t)y]]}}\marginnote{\textcolor{blue}{RC}}}{e^{-i2\pi[\Delta f_0(x,y)t + [k_x(t)x+k_y(t)y]]}} dxdy
\label{eqn:signal_omega_1D_2}
\end{equation}
If \iftoggle{annotated}{\textcolor{blue}{$[k_x (t)x + k_y (t)y] >> \Delta f_0(x,y)t$}\marginnote{\textcolor{blue}{RC}}}{$[k_x (t)x + k_y (t)y] >> \Delta f_0(x,y)t$}, $s(t)$ will accurately encode the object at $m(x,y)$ for k-space location $(k_x(t),k_y(t))$. However, if there is large off-resonance (or long time $t$), the phase pattern associated with the encoding will be warped, as seen in Fig.~\ref{fig:encoding}.

\iftoggle{annotated}{\textcolor{blue}{Figure~\ref{fig:encoding} shows the ideal vs. the actual encoding pattern, i.e., the phase of the complex exponential in Eq.~(\ref{eqn:signal_omega_1D_2}), to acquire the sample at $(k_x,k_y)=(0,0.426 \text{ cm}^{-1})$. Note that Fig.~\ref{fig:encoding} shows the phase pattern superimposed on the object \emph{for a single point in k-space}, and the encoding error is only a function of the time since RF excitation, $t$, and the off-resonance fieldmap, $\Delta f_0(x,y)$, not the sampling trajectory used. The specific geometric distortions in image space will be a function of what these encoding errors were \emph{across all k-space points}, and will depend on the sample ordering, with later samples having more excess phase accrued due to off-resonance. }\marginnote{\textcolor{blue}{RC}}}{Figure~\ref{fig:encoding} shows the ideal vs. the actual encoding pattern, i.e., the phase of the complex exponential in Eq.~(\ref{eqn:signal_omega_1D_2}), to acquire the sample at $(k_x,k_y)=(0,0.426 \text{ cm}^{-1})$. Note that Fig.~\ref{fig:encoding} shows the phase pattern superimposed on the object \emph{for a single point in k-space}, and the encoding error is only a function of the time since RF excitation, $t$, and the off-resonance fieldmap, $\Delta f_0(x,y)$, not the sampling trajectory used. The specific geometric distortions in image space will be a function of what these encoding errors were \emph{across all k-space points}, and will depend on the sample ordering, with later samples having more excess phase accrued due to off-resonance. }
As described in more detail below, for Cartesian sequences this encoding error normally manifests in geometric distortions, and for non- Cartesian sequences, such as spiral, this will result in both distortions and blurring.

%%%%%%%%%%%%%%%%%%%%%%%%%%%%%%%%%%%%%%%%%%%%%%%%%%%%%%
%%%%%%%%  long readouts %%%%%%%%%%%%%%%%%%%
\mysubsection{Utility of long readouts}
In an examination of Eq.~(\ref{eqn:B0phase}), we can see that off-resonance phase accrual is linearly dependent on the time since excitation. This can be split into two parts: the phase accumulation to the echo time (TE) and phase accumulation during k-space scanning.  The first term contributes to signal loss in the $m(x,y)$ term in gradient echo imaging.  The second term contributes to k-space distortions and is effects, as shown in Fig.~\ref{fig:encoding}.  Thus, long readouts for imaging will show increased distortions.  That said, there are advantages to having longer readouts in terms of SNR, $k$-space scanning efficiency, and overall acquisition speed.
\mysubsubsection{SNR}
For a given repetition time (TR), flip angle, and spatial resolution, the signal to noise ratio (SNR) is proportional to the square root of the total acquisition time, $T_{A/D}$~\cite{Macovski1996NoiseMRI}, at least to a first approximation. The total acquisition time here is defined the sum of the length of all acquisitions contributing to a given image. One can also define SNR efficiency as the SNR divided by the square root of overall time to acquire an image. Thus, as one increases the time spent acquiring data in each TR, 
\iftoggle{annotated}{\textcolor{blue}{say by filling in dead times, the SNR efficiency also increases. This is, however, often a minor effect.}\marginnote{\textcolor{blue}{C}}}{say by filling in dead times, the SNR efficiency also increases. This is, however, often a minor effect.}

\mysubsubsection{{k}-space scanning}
\iftoggle{annotated}{\textcolor{blue}{Perhaps the most important advantage of a long readout is for speeding up the image acquisition process, by acquiring more k-space data with each RF excitation.}\marginnote{\textcolor{blue}{RC}}}{Perhaps the most important advantage of a long readout is for speeding up the image acquisition process, by acquiring more k-space data with each RF excitation.}
In general, reduced study duration is beneficial for patient comfort and for economic reasons. Beyond that, accelerated imaging is also enabling for many applications.  While there are numerous methods to accelerate imaging, notably parallel imaging, k-space scanning with long readouts also plays an important role in reducing scan time.  For example, in dynamic imaging applications like cardiac imaging, there may be advantages to acquiring a larger part of k-space with each excitation so that higher resolution images can be acquired during a breath hold. 
\iftoggle{annotated}{\textcolor{blue}{In neuroimaging, the ability to ``freeze" motion has led to single shot k-space scanning, most commonly using echo planar imaging (EPI), to be the dominant methods for acquiring diffusion weighted and BOLD functional brain images.}\marginnote{\textcolor{blue}{RC}}}{In neuroimaging, the ability to ``freeze" motion has led to single shot k-space scanning, most commonly using echo planar imaging (EPI), to be the dominant methods for acquiring diffusion weighted and BOLD functional brain images.}
These are but a few examples where long readout and k-space scanning are useful in MRI.  

%%%%%%%%%%%%%%%%%% figure %%%%%%%%%%%%%%%%%%%%%

%%%%%%%%%%%%%%%%%%%%%%%%%%%%%%%%%%%%%%%%%%
%%%%%%%%  EPI, Cartesian, etc. %%%%%%%%%%%%%
\mysubsection{Cartesian imaging: spin-warp and EPI}

In Cartesian imaging, data is sampled along straight line segments along $k_x$.
The term ``Cartesian'' implies that the samples are regularly spaced, though in practice the $k_x$ dimension is sometimes sampled along the trapezoidal gradient ramps (in addition to the plateau) for increased scan efficiency. 
Since the invention of MRI, Cartesian imaging has remained the ``bread and butter" MRI technique despite its relatively low scan-time efficiency compared to non-Cartesian imaging, since 
(i) it allows fast and simple reconstruction using inverse Fast Fourier Transforms (iFFT), 
(ii) image artifacts due to off-resonance and chemical shift are often easy to recognize and ``read through'', and
(iii) geometric distortions can be corrected using fast and widely available methods.

\mysubsubsection{Spin-warp imaging}
To understand the impact of off-resonance and chemical shift on Cartesian imaging, consider first the case of ``spin-warp'' imaging, where a single line along $k_x$ is sampled after each RF excitation, with constant gradient amplitude $G_x$.
In this case the $k_x$-encoding term in Eq.~(\ref{eqn:signal_omega_1D_2}) can be written as 
$k_x(t) = \frac{\gamma}{2\pi} G_x t$
and (since the acquisition bandwidth along $k_y$ is effectively infinite) the exponent in Eq.~(\ref{eqn:signal_omega_1D_2}) becomes
$-i2\pi k_x [x + \delta x] - i2\pi k_y(t)y$ where
\begin{equation}
\delta x = \frac{
\iftoggle{annotated}{\textcolor{blue}{2 \pi \Delta f_0(x,y)}\marginnote{\textcolor{blue}{RC}}}{2 \pi \Delta f_0(x,y)}
}{\gamma G_x}.
\label{eqn:dx}
\end{equation}
In other words, a voxel at position $x$ with off-resonance frequency 
\iftoggle{annotated}{\textcolor{blue}{$\Delta f_0$}\marginnote{\textcolor{blue}{RC}}}{$\Delta f_0$}
will appear at $x + \delta x$ in the reconstructed image.
For typical readout bandwidths, $\delta x$ is negligible for most tissue types at clinical field strengths, but can be on the order of a voxel width, e.g., fat.
As a consequence, the acquisition bandwidth for most clinical sequences is kept sufficiently high to avoid significant fat shift.
\iftoggle{annotated}{\textcolor{blue}{For example, a common acquisition bandwidth is +/- 32 kHz (64 kHz total). With the fat chemical shift at 3T (448 Hz), the corresponding image shift it 1.75 voxels for a 256 readout sample.}\marginnote{\textcolor{blue}{RC}}}{For example, a common acquisition bandwidth is +/- 32 kHz (64 kHz total). With the fat chemical shift at 3T (448 Hz), the corresponding image shift it 1.75 voxels for a 256 readout sample.}
Note that Eq.~(\ref{eqn:dx}) ignores any $B_0$ inhomogeneity \textit{within} the voxel centered at location $(x,y)$, and hence does not describe any signal loss that may occur at a given echo time (TE).

\mysubsubsection{Echo-planar imaging}
In echo-planar imaging (EPI), 
\iftoggle{annotated}{\textcolor{blue}{multiple lines at different $k_y$-encoding levels are acquired after RF excitation.}\marginnote{\textcolor{blue}{RC}}}{multiple lines at different $k_y$-encoding levels are acquired after RF excitation.}
This is done by playing a small y-gradient trapezoid ``blip'' after the end of each $k_x$ line, and then traversing the subsequent line in reverse direction (non-flyback EPI).
Assuming regular sampling along $k_y$ with subsequent ``echoes'' spaced $\tau_s$ apart in time, and ignoring the (typically small) shift $\delta x$ along the frequency-encoding direction, the exponent in Eq.~(\ref{eqn:signal_omega_1D_2}) becomes $-i2\pi k_x x - i2\pi k_y [y + \delta y]$ where
\begin{equation}
\delta y = \frac{
\iftoggle{annotated}{\textcolor{blue}{2 \pi \Delta f_0(x,y)}\marginnote{\textcolor{blue}{RC}}}{2 \pi \Delta f_0(x,y)}
}{\gamma \overline{G_y}} 
\label{eqn:dy}
\end{equation}
is the spatial shift (along $y$) in the reconstructed image.
Here, $\overline{G_y}$ is the \textit{average} gradient amplitude of the $y$ blip over the echo time spacing $\tau_s$, which is much smaller than the readout gradient amplitude $G_x$ in Eq.~(\ref{eqn:dx}).
\iftoggle{annotated}{\textcolor{blue}{As a result, EPI can produce substantial spatial shifts along the phase encoding (PE) direction, here $k_y$. However if the user were to change the PE direction, the spatial shifts would also change direction, as these off-resonance induced spatial shifts are primarily in the PE direction.}\marginnote{\textcolor{blue}{RC}}}{As a result, EPI can produce substantial spatial shifts along the phase encoding (PE) direction, here $k_y$. However if the user were to change the PE direction, the spatial shifts would also change direction, as these off-resonance induced spatial shifts are primarily in the PE direction.}

\iftoggle{annotated}{\textcolor{blue}{A useful way to think about these EPI distortions is to express the average gradient in the PE direction in units of Hz/cm, as follows:
\begin{equation}
    \overline{G}_{y, Hz/cm} = \frac{\gamma \overline{G_y}}{2 \pi} = 1/(\tau_s*FOV/R)
\label{eqn:dy2}
\end{equation}
where $\tau_s$ = echo spacing and R = in-plane acceleration. Typical values are about 66 Hz/cm for high-resolution fMRI (2mm isotropic resolution, $\tau_s$ = 0.75 ms, FOV = 20 cm, R = 1), 100 Hz/cm for low-resolution fMRI ($\tau_s$ = 0.50 ms, FOV = 20 cm, R = 1), and about 133 Hz/cm for DTI with 2mm resolution and in-plane acceleration ($\tau_s$ = 0.75 ms, FOV = 20 cm, R = 2).
The spatial shift along the PE direction (Eq.~(\ref{eqn:dy})) is then found by dividing the local off-resonance $\Delta f_0$ by the average gradient expressed in Hz/cm.}}{A useful way to think about these EPI distortions is to express the average gradient in the PE direction in units of Hz/cm, as follows:
\begin{equation}
    \overline{G}_{y, Hz/cm} = \frac{\gamma \overline{G_y}}{2 \pi} = 1/(\tau_s*FOV/R)
\label{eqn:dy2}
\end{equation}
where $\tau_s$ = echo spacing and R = in-plane acceleration. Typical values are about 66 Hz/cm for high-resolution fMRI (2mm isotropic resolution, $\tau_s$ = 0.75 ms, FOV = 20 cm, R = 1), 100 Hz/cm for low-resolution fMRI ($\tau_s$ = 0.50 ms, FOV = 20 cm, R = 1), and about 133 Hz/cm for DTI with 2mm resolution and in-plane acceleration ($\tau_s$ = 0.75 ms, FOV = 20 cm, R = 2).
The spatial shift along the PE direction (Eq.~(\ref{eqn:dy})) is then found by dividing the local off-resonance $\Delta f_0$ by the average gradient expressed in Hz/cm.}
\iftoggle{annotated}{\textcolor{blue}{}\marginnote{\textcolor{blue}{RC}}}{}

\mysubsubsection{Interleaved and readout mosaic segmented EPI}
EPI comes in several multi-shot variants that make various tradeoffs between image artifacts and scan efficiency.
In addition to geometric distortions as just described, another consideration is $T_2^*$ decay during the echo train, which acts as a k-space filter resulting in blurring in image space.

In interleaved EPI, the area of each $k_y$-encoding blip, and hence $\overline{G_y}$, is increased, making each EPI train undersampled by a factor equal to the number of segments (shots) $N_s$.
This reduces the geometric shifts, and the time allowing for $T_2^*$ decay, by that same factor $N_s$.
Interleaved EPI is typically implemented by shifting the start of subsequent EPI segments by a time $n/\tau_s$ where $n = 0, ..., N_s-1$ is the shot number, to avoid abrupt phase jumps in $k_y$-space due to off-resonance.
A potential drawback of interleaved EPI is shot-to-shot variability in the acquired data, resulting from, e.g., physiological fluctuations or subject motion.

An alternative scheme useful for diffusion MRI is readout mosaic segmented EPI, where only a portion (``blind'') of $k_x$-space is acquired during each shot, while sampling all desired $k_y$ locations per shot~\cite{Porter2004Multi-shotCorrection,Holdsworth2008Readout-segmented3T}.
The advantage of this approach is that it preserves the regular sampling along $k_y$ in the presence of shot-to-shot k-space shifts due to rotational motion in the presence of diffusion-encoding gradients~\cite{Butts1996Diffusion-weightedEchoes}.
By introducing a small overlap between blinds along $k_x$, such shifts need not introduce any gaps along $k_x$ as well.
A drawback of this approach is that the echo time spacing is not reduced by the full segmentation factor $N_s$ due to finite gradient rise time, and hence the reduction in geometric shift is somewhat smaller than for interleaved EPI for a given $N_s$.

%%%%%%%%%%%%%%%%%%%%%%%%%%%%%%%%%%%%%%%%%%%%%%%%%
%%%%%%%%      non-Cartesian   %%%%%%%%%%%%%
\mysubsection{Non-Cartesian acquisitions}
While much of MRI carried out on clinical scanners uses Cartesian acquisitions like spin-warp or EPI, there are numerous useful applications that use non-Cartesian acquisitions.  This is a broad class of image acquisitions that typically do not acquire data along parallel lines in k-space. Some aspects are similar to Cartesian imaging, for example, the need to acquire data over a prescribed area in k-space for the desired spatial resolution and to acquire k-space data with sufficient sample density to prevent aliasing. The nature of the off-resonance artifacts, however, are quite different.  Below, we discuss these effects for radial line and spiral imaging, but note that there are yet other non-Cartesian trajectories, e.g., rosettes~\cite{Noll1997MultishotImaging}, that have quite complicated off-resonance behavior that is beyond the scope of this review.

\mysubsubsection{Radial line and spiral imaging}
Radial line and spiral imaging are useful in a number of applications.  For example, radial line acquisitions in 2D or 3D have substantial utility in zero- and ultra-short TE (ZTE and UTE) imaging~\cite{Weiger2013ZTEHumans,Gatehouse2003MagneticTissue}, angiography~\cite{Du2004Contrast-enhancedReconstruction}, and contrast enhanced abdominal imaging~\cite{Feng2014Golden-angleMRI}.  Spiral imaging has been shown to be useful in functional MRI~\cite{Noll1995SpiralActivation}, cardiac imaging~\cite{Nayak2005SpiralImaging}, and brain imaging~\cite{Li2020ImprovingTechnique}.

Spiral and radial line imaging have off-resonance features that are similar to each other, but are quite different from spin-warp and EPI. For each line of a radial line acquisition, there is a shift in the radial direction in the form:
\begin{equation}
\delta r \equiv \frac{
\iftoggle{annotated}{\textcolor{blue}{2 \pi \Delta f_0(x,y)}\marginnote{\textcolor{blue}{RC}}}{2 \pi \Delta f_0(x,y)}
}{\gamma G_r} 
\label{eqn:dr}
\end{equation}
where $G_r$ is the strength of the radial gradient.  Since the direction of the radial gradient is constantly changing and ultimately points in all directions, the net effect is not a single shift, but a shift in all directions.  More specifically, any off-resonant point in the object maps to a ring-like response of radius $\delta r$. This is equivalent to saying that the point spread function (PSF) or impulse response is this ring-like function, the radius of which, $\delta r$ is proportional to the off-resonance frequency, 
\iftoggle{annotated}{\textcolor{blue}{$\Delta f_0(x,y)$}\marginnote{\textcolor{blue}{RC}}}{$\Delta f_0(x,y)$}
. For a complex image, this manifests as a blur of radius $\delta r$. 

For spirals, the off-resonance response is again a radial shift in all directions, with amplitude:
\begin{equation}
\delta r_s \approx \frac{
\iftoggle{annotated}{\textcolor{blue}{2 \pi \Delta f_0(x,y)}\marginnote{\textcolor{blue}{RC}}}{2 \pi \Delta f_0(x,y)}
}{\gamma \overline{G_r}} 
\label{eqn:dspiral}
\end{equation}
where $\overline{G_r}$ is the average outward component of the spiral gradients.  One can also infer the average outward component with $\overline{G_r} = 2 \pi k_{max} / T$, where $T$ is the length of the spiral readout. Now, any off-resonant point in the object maps to a ring-like response of radius $\delta r_s$ and as with radial imaging, this is proportional to the off-resonance frequency, 
\iftoggle{annotated}{\textcolor{blue}{$ \Delta f_0(x,y)$}\marginnote{\textcolor{blue}{RC}}}{$ \Delta f_0(x,y)$}
.  As can be seen from the alternate $\overline{G_r}$ expression, the blur is also proportional to length of the spiral readout, $T$.  Again, for a complex image, this manifests as a blur of radius $\delta r_s$.  For spiral imaging, the relationship that the blur is a ring of radius $\delta r_s$ is only approximate because the average outward component typically varies during the readout making the actual blur function a bit more complicated.  Still, the main principle is that longer readouts make the off-resonance image blurring worse.

%%%%%%%%%%%%%%%%%%%% figure 
\begin{figure}[h]
\centering
\includegraphics[width=\textwidth]{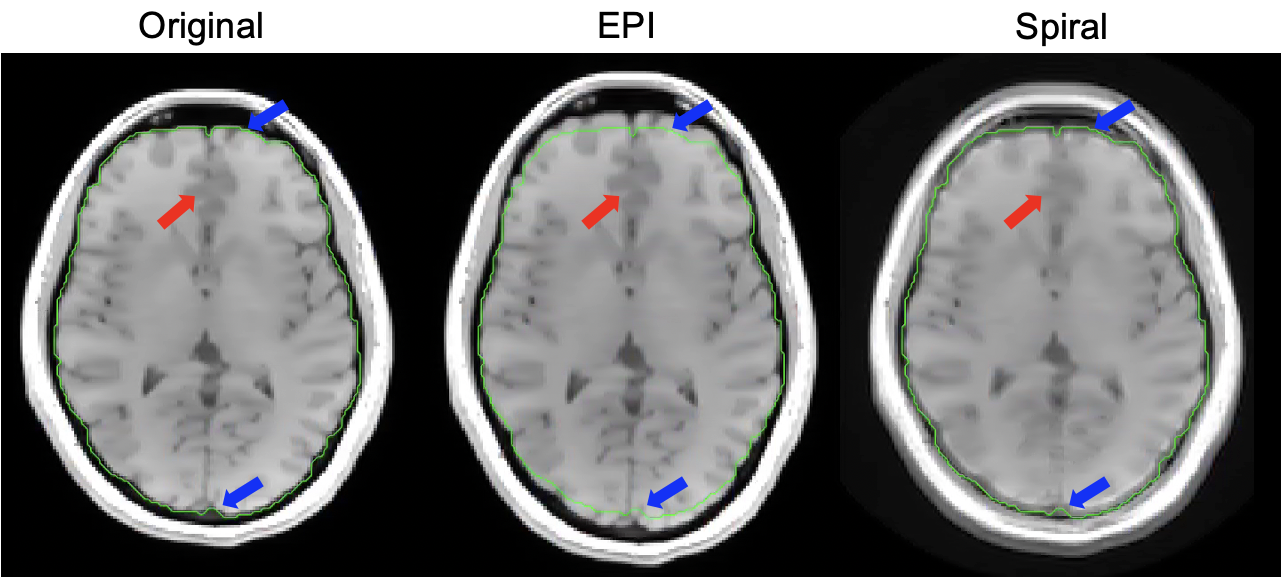}
\iftoggle{annotated}{\marginnote{\textcolor{blue}{RC}}}{}
\caption{\textbf{Image simulations: Cartesian vs. non-Cartesian.}
\footnotesize{Here we show the different off-resonance effects from EPI vs. spiral acquisition. In the EPI image, geometric distortion of the brain is visible \iftoggle{annotated}{\textcolor{blue}{with a distortion sensitivity of 75 Hz/cm (see Eq.~(\ref{eqn:dy2})), a typical value,}}{with a distortion sensitivity of 75 Hz/cm (see Eq.~(\ref{eqn:dy2})), a typical value,} resulting in the brain being stretched compared to the original}
\iftoggle{annotated}{\footnotesize{\textcolor{blue}{Cartesian spin-warp image}}}{\footnotesize{ Cartesian spin-warp image}}\footnotesize{ (blue arrows). In the spiral image, there is a large amount of blurring, noted by the red arrow, but there is less geometric distortion than in EPI. }} 

\label{fig:image_sims}
\end{figure}

The off-resonance behavior also extends to three dimensions.  For examples, ZTE imaging and angiography applications often use a 3D radial line acquisition.  While spirals don't extend directly to 3D, there are 3D trajectories like 3D cones, \iftoggle{annotated}{\textcolor{blue}{trajectories that spiral outward while rotating along two axes (sometimes called yarnball trajectories),}\marginnote{\textcolor{blue}{RC}}}{trajectories that spiral outward while rotating along two axes (sometimes called yarn trajectories),}
 and spiral-radial hybrids~\cite{Irarrazabal1995FastImaging} that have off-resonance properties similar to spirals.  In these cases, we consider induced blur to be occurring in all three dimensions, that is, the PSF or impulse response is an annular sphere with radius $\delta r$ (or $\delta r_s$).  Again, greater off-resonance and longer readouts increase the radius of the blurring function.

The ways that off-resonance affects EPI and spiral images at 3T is shown in Fig.~\ref{fig:image_sims} (images in Fig.~\ref{fig:image_sims} were acquired from a healthy subject after obtaining IRB approved written informed consent). As expected the artifacts in the spiral image are primarily blurring, and the artifacts in the simulated EPI image are primarily geometric distortions along the phase encode (vertical) direction. 
 \iftoggle{annotated}{\textcolor{blue}{Additional details on the simulation can be found at \url{https://github.com/fmrilab/B0-review-2022}.}\marginnote{\textcolor{blue}{RC}}}{Additional details on the simulation can be found at \url{https://github.com/fmrilab/B0-review-2022}.}

%%%%%%%%%%%%%%%%%%%%%%%%%%%%%%%%%%%%%%%%%%%%%%%%%%%%%%%%
%%%%%%%%  slice shifting, RF bandwidth  %%%%%%%%%%%%%
\mysubsection{Imaging near metal implants}

Though not ferromagnetic, the susceptibility differences between metals and tissues are substantially larger than the differences between tissues and the surrounding air. These large susceptibility differences, in combination with the often large mass of these implants, lead to similar, but magnified, $B_0$ effects.  For example, much of GRE imaging is infeasible in close proximity to implants, except for the very shortest TEs, meaning that most studies use SE or Fast/Turbo Spin Echo (FSE/TSE) acquisitions.  Similarly, rapid imaging methods like EPI or spiral imaging with implants will typically produce artifacts that are too severe to correct.  There are often severe geometric distortions in the readout (frequency encoding) direction associated with implants, including signal voids and piling up artifacts.  Another common issue is the distortion of the slice profiles, where there can be spatial displacement several slice thicknesses in magnitude, distorted slice profiles like potato chips, and variable slice thicknesses.  In both frequency encode and slice effects, the spatial distortions are proportional to the frequency offset divided by the bandwidth per pixel (or slice thickness).

\section{Estimating \texorpdfstring{\emph{B}\textsubscript{0}}{} Field Inhomogeneity} \label{sec:measuringB0}

%%%%%%%%%%%%%%%%%%%%%%%%%%%%%%%%%%%%%%%%%%%
%%%%%%   sequences based approaches
\mysubsection{MRI sequences for acquiring fieldmaps}
Before correction for $B_0$ inhomogeneity artifacts, it is often necessary to acquire and estimate the fieldmap. The simplest way to estimate a $B_0$ fieldmap is to create a pulse sequence that acquires two sequential gradient echo images at different TEs, and then subtract their relative phases to obtain the amount of precession at that voxel over the echo time difference $\Delta$TE. The phase difference is then scaled by $\Delta$TE to get the fieldmap:
\begin{equation}
    \Delta \phi (x,y) = \angle m(x,y,\mathrm{TE}_1) - \angle m(x,y,\mathrm{TE}_2)
\end{equation}
\begin{equation}
    \Delta \omega_0(x,y) = \frac{\Delta \phi (x,y) }{\Delta\mathrm{TE}}
    \label{eqn:delta_omega}
\end{equation}
This is the basic idea behind $B_0$ fieldmapping in MRI, but there are many ways to estimate a fieldmap within this general approach. The choice of $\Delta$TE will affect the quality of the fieldmaps, and has been quantitatively evaluated in prior work~\cite{Hutton2002ImageEvaluation}. Normally gradient echo images at short TEs are used to avoid distortions in the underlying fieldmap, but for applications using EPI, there has also been work in using EPI-based fieldmaps~\cite{Reber1998CorrectionMaps}. While EPI fieldmaps will be distorted given the longer TEs required, they have the advantage that they will be distorted in the same way and also experience the same eddy currents as the EPI data used in the imaging experiment, potentially allowing for easier look up of the off-resonance at any voxel.

Another consideration when calculating fieldmaps is phase wrap. A longer $\Delta$TE is desirable because it will increase the SNR of the phase difference measurement. However, when $\Delta$TE is increased, more phase wrap can occur. Fortunately, many tools such as FSL~\cite{Jenkinson2012FSL} include phase unwrapping algorithms that are useful when calculating off-resonance field maps, such as the automated method introduced by Jenkinson in 2003~\cite{Jenkinson2003FastAlgorithm}.

There have also been recent approaches to rapidly acquire $B_0$ fieldmaps along with transmit coil patterns ($B_1^{+}$), receive coil sensitivity ($B_1^{-}$), and eddy current maps in as little as 11s at 1$\times$2$\times$2mm$^{3}$ resolution at 3T~\cite{Iyer2020PhysiCal:Mapping}. This kind of approach is especially useful for model-based image reconstructions that incorporate $B_0$ (see Section \ref{subsec:mbir}), since $B_1^{-}$ maps are often needed as well.

\iftoggle{annotated}{\textcolor{blue}{There are requirements that the fieldmaps be reasonable accurate and not too noisy.  As a general rule, the error/noise level should be below the level of observable degradation of the point spread function. One rule of thumb is for the error and noise levels to be below 0.25 cycle of phase over the readout.  For example, if you want to correct the field effects for a 25 ms readout, 40 Hz offset will produce one cycle of phase over the readout and thus, we would want worst case errors to be below 10 Hz.}\marginnote{\textcolor{blue}{RC}}}{There are requirements that the fieldmaps be reasonable accurate and not too noisy.  As a general rule, the error/noise level should be below the level of observable degradation of the point spread function. One rule of thumb is for the error and noise levels to be below 0.25 cycle of phase over the readout.  For example, if you want to correct the field effects for a 25 ms readout, 40 Hz offset will produce one cycle of phase over the readout and thus, we would want worst case errors to be below 10 Hz.}
\iftoggle{annotated}{\textcolor{blue}{Often times, the fieldmaps are smoothly varying except for small air/tissue boundaries.  As such, it is often expedient to collect fieldmaps at lower spatial resolution (2-3 mm) than the images to be corrected (1-2 mm).
}\marginnote{\textcolor{blue}{RC}}}{Often times, the fieldmaps are smoothly varying exec pt for small air/tissue boundaries.  As such, it is often expedient to collect fieldmaps at lower spatial resolution (2-3 mm) than the images to be corrected (1-2 mm).}

%%%%%%%%%%%%%%%%%%%%%%%%%%%%%%%%%%%%%%%%%%%%%%%%%%%%%%%%%
%%%%%%%%%   Hardware
\mysubsection{Hardware methods}
In addition to measuring a fieldmap using a sequence, it is possible to employ external hardware and/or built in scanner hardware to monitor changes in off-resonance. Generally these types of hardware tracking methods will be used for dynamic $B_0$ correction (see Section \ref{subsec:dynam}) after an initial $B_0$ map has been measured using a fieldmapping sequence. One option is to use NMR field probes~\cite{Barmet2008SpatiotemporalMR,Dietrich2016AAnalysis,Gross2016DynamicResolution}, which measure the changes in phase of the MRI signal within small external samples. These probes are now commercially available (\url{https://skope.swiss/}) and have been used in applications such as diffusion MRI, fMRI, and ultra high field imaging. Another approach is to use FID measurements from a multicoil receive array, which can encode changes in signal based on the different spatial locations of the coils. By combining with a reference image from each coil, Wallace et al. showed that FIDs can be used to dynamically track up to 2$^\text{nd}$ order field changes~\cite{Wallace2020RapidImage}.

%%%%%%%%%%%%%%%%%%%%%%%%%%%%%%%%%%%%%%%%%%%%%%%%%%%%%%%%%
%%%%%%%%%%% iterative
\mysubsection{Optimization and smoothing methods}

After multi-TE data has been acquired, Eq.~(\ref{eqn:delta_omega}) is the simplest way to calculate a fieldmap, but the maps are often improved by using some form of smoothing or regularization due to the difficulty of getting $B_0$ values near the edges of anatomy. In the original TOPUP paper~\cite{Jezzard1995CorrectionVariations}, a smoothed version of the fieldmap that is based on polynomial fitting was used. The work by Hutton et al.~\cite{Hutton2002ImageEvaluation}, in addition to examining $\Delta$TE, also quantitatively evaluates how the amount of smoothing used affects fieldmap estimates. 

There are also optimization-based approaches that use a multi-term objective function, where the phase differences between multiple TEs is a data fit term and another regularization term penalizes rapidly varying and sudden field changes~\cite{Funai2008RegularizedMRI}. This method has been expanded upon to incorporate a more efficient optimization algorithm and the ability to input multicoil data~\cite{Lin2020EfficientMRI}. Additionally, it is possible to jointly optimize for a fieldmap and an image, and these approaches are described in Section \ref{subsec:mbir} on model based image reconstruction.

\mysection{Correction Strategies}\label{sec:corrections}

In this section we discuss strategies to correct for off-resonance artifacts in MRI. Many of the methods described could fit into multiple categories, but we have broadly categorized them to help in describing the different options available to MRI users, as outlined in Figure \ref{tab:corrs}. 

\begin{figure}[H]
\begin{tabular}{ | m{7cm} | m{10cm}|} 
  \hline
  \textbf{Artifact Prevention (Section \ref{subsec:artifact_reduction})} & \textbf{$\boldsymbol{B_0}$ Correction Methods (Sections \ref{subsec:cart_corr}-\ref{subsec:slice_shift_corr})}  \\ 
  \hline \begin{itemize}[leftmargin=*]
    \item Hardware based
        \begin{itemize}
         \item passive \& active shimming
         \item lower-field MRI
        \end{itemize}
    \item Sequence based
        \begin{itemize}
            \item spin echoes      
            \item shorter readouts: parallel imaging, multishot imaging
            \item fat suppression / nulling (STIR)
            \item smaller voxels: parallel imaging, simultaneous multislice
        \end{itemize}
  \end{itemize} & \begin{itemize}[leftmargin=*]
  
      \item EPI/Cartesian focused 
      \begin{itemize}
        \item Image Space: TOPUP
        \item PSF
      \end{itemize}
      
      \item Non-Cartesian focused 
      \begin{itemize}
        \item Direct/Analytical: conjugate phase, SPHERE
        \item Autofocusing
         \end{itemize}
        \item Model based image reconstruction including $B_0$
      \item Dynamic corrections
      \item Corrections for larger $B_0$ shifts around metal implants
      \begin{itemize}[nosep]
        \item View angle tilting
        \item MAVRIC
        \item SEMAC      \vspace{-12pt}
      \end{itemize} 
    \end{itemize} \\ \hline
\end{tabular}
\caption{\textbf{Common $B_0$ artifact mitigation strategies.} } 
\label{tab:corrs}
\end{figure}

%%%%%%%%%%%%%%%%%%%%%%%%%%%%%%%%%%%%%%%%%%%%%
%%%% prevention %%%%%%%%%%%%%%%%%%%%%%%%%
\vspace{-17pt}
\mysubsection{Artifact Prevention \& Reduction}\label{subsec:artifact_reduction}
\myspace
\mysubsubsection{Hardware based artifact mitigation}
The most straightforward way to prevent $B_0$ artifacts is to have as uniform a field as possible before starting a scan, and the process of refining the uniformity of the $B_0$ field is referred to as \emph{shimming}~\cite{Schneider1991RapidShimming}. Shimming can be done either by improving the static $B_0$ field hardware (i.e., with superconducting shim magnets or ferrous passive elements) or by actively changing the currents in the scanner's electro-shim coils, which include the gradient coils. An example of passive shimming at low field is the work of Liu, et al., where on their 50mT scanner they originally had a 2000 ppm peak-to-peak uniformity over a 24cm diameter spherical volume (DSV) within their SmCo dipole magnet, but after adding smaller passive shim magnets their uniformity was increased to 250 ppm~\cite{Liu2021AScanner}. Automated shimming is a standard part of a clinical MRI pre-scan calibration (at least for the constant linear shims terms) and for most users this will be automatic, but for certain types of acquisitions, such as MR spectroscopy that requires a very uniform field~\cite{Juchem2021BRecommendations}, the user may need to do  manual shimming as well. 

In addition to shimming, using a lower magnetic field is a direct way to reduce $B_0$ artifacts. As was discussed in Section \ref{sec:sources}, $B_0$ field changes from susceptibility and chemical shift scale with field strength, so these artifacts are generally worse at higher field. There are some $B_0$ artifact mitigation strategies that are harder at low field due to decreased SNR, but in general going to lower field will reduce signal loss and encoding artifacts from an inhomogeneous field.

\mysubsubsection{Sequence based artifact mitigation}
When designing an MRI sequence, parameters can be adjusted to reduced the effects of off-resonance. Using a short readout time (high acquisition bandwidth) will limit the amount of extra phase accrued due to off-resonance and therefore reduce artifacts. The use of parallel imaging techniques~\cite{Larkman2004, Larkman2007a} is an effective way to reduce the readout length of \iftoggle{annotated}{\textcolor{blue}{a single-shot acquisition (such as EPI or single-shot spiral)}\marginnote{\textcolor{blue}{RC}}}{a single-shot acquisition (such as EPI or single-shot spiral)}, by allowing the user to collect less k-space data each shot. If using longer readouts, spin echoes and spin echo trains will refocus spins and avoid much of the $T_2^*$ signal loss~\cite{Hahn1950SpinEchoes,Hennig1986}. Chemical shift artifacts can be avoided by using a short time inversion recovery (STIR) sequence that nulls recovering fat signal, inserting fat saturation modules before water excitation (an RF excitation at the fat frequency and then a large crusher gradient), or by using a spectrally selective RF pulse centered around the water resonant frequency~\cite{Bley2010FatImaging}. Further, while this generally has the downside of longer scan time due to additional encoding, using smaller voxels will decrease the possible signal loss due to a more local neighborhood of spins being averaged in each voxel. Parallel imaging and simultaneous multislice imaging~\cite{Feinberg2013Ultra-fastImaging} are effective strategies to reduce voxel size without increasing scan time.

%%%%%%%%%%%%%%%%%%%%%%%%%%%%%%%%%%%%%%%%%%%%%%%%%%%%
%%%%  EPI/Cartesian  %%%%%%%%%%%%%%%%%%%%%%%%%
\mysubsection{EPI}\label{subsec:cart_corr}

%%%%%%%%%%%%%%%%%%%  figure  %%%%%%%%%%%%%%%%%%%%
\begin{wrapfigure}{R}{.43\textwidth}
\centering
\includegraphics[width=.43\textwidth]{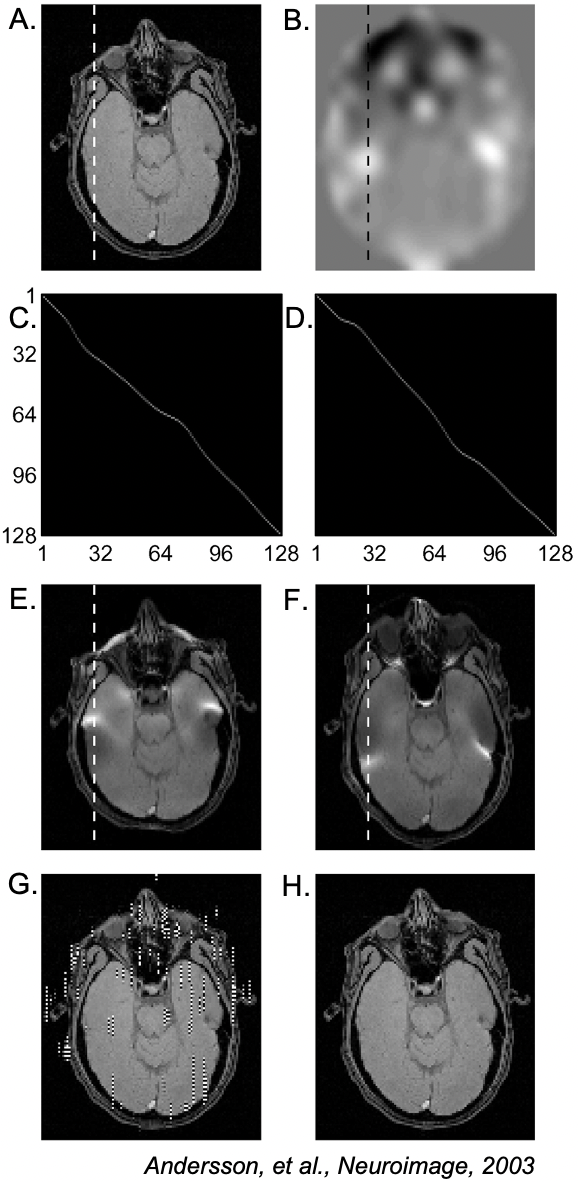}
\caption{ \footnotesize{ 
\textbf{Illustration of EPI distortion correction using Eq.~(\ref{eqn:topup}).}
(A) Ground-truth (undistorted) object and (B) fieldmap.
(C,D) $\boldsymbol{K}_{+,i}$ and $\boldsymbol{K}_{-,i}$ for the dashed lines in (E,F).
(E,F) Distorted images $\boldsymbol{f}_+$ and $\boldsymbol{f}_-$.
(G) Distortion correction attempt using only $\boldsymbol{f}_+$.
(H) Successful distortion correction using Eq.~(\ref{eqn:topup}).
\emph{(Images reused from Andersson et al., 2003~\cite{Andersson2003HowImaging} with permission.)}}
} 
\label{fig:topup}
\end{wrapfigure}

Several methods for correcting geometric and intensity distortions in EPI have been proposed (see, e.g.,~\cite{Schallmo2021AssessingData, Abreu2021QuantitativeAnalyses} for recent overviews), and here we focus on three of the more popular ones:
image-based correction using an acquired $B_0$ map, image-based correction using two reference scans with opposite phase-encode blip directions, and the point-spread function (PSF) mapping approach.
These have all proven to be useful for EPI distortion correction in fMRI and diffusion imaging.

\mysubsubsection{Fieldmap based correction} \label{subsubsec:fm_corr}
One popular approach is based on the separate acquisition of an undistorted $B_0$ map (e.g., using spin-warp 3D GRE), that is used to calculate the shift $\delta y$ (Eq.~(\ref{eqn:dy})) for each spatial location~\cite{Jezzard1995CorrectionVariations}.  
The pixel values in the reconstructed (distorted) image are then shifted along the phase-encode direction by $-\delta y$.
This is followed by an intensity correction step based on the spatial gradient of $\delta y$~\cite{Jezzard1995CorrectionVariations}.
Potential drawbacks of this approach include inaccurate correction near object edges due to inaccurate $B_0$ estimates (partial-volume effects),
and inability to resolve overlapping pixels.

\mysubsubsection{Opposite phase-encoding scans} 
Another widely used approach that also operates on magnitude images is TOPUP~\cite{Andersson2003HowImaging}, that uses as input two EPI images that differ only in the sign of the phase-encoding blips (i.e., traversing $k_y$ in the positive and negative, or bottom-up and top-down, directions).
The TOPUP model is 
\begin{equation}
 \begin{bmatrix}
 \boldsymbol{f}_{+,i} \\
 \boldsymbol{f}_{-,i}
 \end{bmatrix}
 = 
 \begin{bmatrix}
 \boldsymbol{K}_{+,i} \\
 \boldsymbol{K}_{-,i}
 \end{bmatrix}
 \boldsymbol{x}_i
 \label{eqn:topup}
\end{equation}
where $\boldsymbol{x}_i$ is the $i^\textrm{th}$ column (along y) of the undistorted true image $\boldsymbol{x}$,
$\boldsymbol{f}_+$ and $\boldsymbol{f}_-$ are the distorted bottom-up and top-down images, respectively, and
$\boldsymbol{K}_{+,i}$ and $\boldsymbol{K}_{-,i}$ are real-valued square matrices that map the pixel locations in the true image column $\boldsymbol{x}_i$ to the observed image columns $\boldsymbol{f}_{+/-,i}$. In the absence of $B_0$ inhomogeneity, $\boldsymbol{K}_{+,i}$ and $\boldsymbol{K}_{-,i}$ would be the identity matrix $\boldsymbol{I}$.

With knowledge of $B_0$ (which determines $\boldsymbol{K}_{+,i}$ and $\boldsymbol{K}_{-,i}$), Eq.~(\ref{eqn:topup}) can be inverted to obtain a least-squares estimate $\boldsymbol{\hat{x}}$ of the true image, with the benefit that pixels that overlap in one of the images (causing either $\boldsymbol{K}_{+,i}$ or $\boldsymbol{K}_{-,i}$ to be singular) are separated in the other, which enables the underlying true image to be resolved.

In the absence of $B_0$ inhomogeneity, $\boldsymbol{\hat{x}}_i$ is simply the average of $\boldsymbol{f}_{+,i}$ and $\boldsymbol{f}_{-,i}$.
Furthermore, the TOPUP model can be used to estimate $B_0$ from the two EPI scans (obviating the need for a separately acquired $B_0$ map), by enforcing consistency between the observed images $\boldsymbol{f}_{+,-}$ and the predicted images $\boldsymbol{\hat{f}}_{+,-}$ created by distorting $\boldsymbol{\hat{x}}$ using Eq.~(\ref{eqn:topup}):
\begin{equation}
    \boldsymbol{\hat{{b}_i}} = \mathop{\mathrm{arg\,min}}_{\boldsymbol{b}_i}
    \,
    \biggl| 
    \begin{bmatrix}
    \mathbf{f}_{+,i} \\
    \mathbf{f}_{-,i}
    \end{bmatrix}
    -
    \begin{bmatrix}
    \mathbf{K}_{+,i}(\boldsymbol{b}_i) \\
    \mathbf{K}_{-,i}(\boldsymbol{b}_i)
    \end{bmatrix}
    \boldsymbol{\hat{x}}_i(\boldsymbol{b}_i)
    \biggr|_2^2
    \label{eqn:topup_b}
\end{equation}
where $\boldsymbol{b}_i$ parametrizes the displacements (using the notation from~\cite{Andersson2003HowImaging}),
and we have made the dependence of $\boldsymbol{\hat{x}}_i$ on $\boldsymbol{b}_i$ explicit.
Knowledge of $\boldsymbol{b}_i$ yields an estimate of the fieldmap $B_0$, which can then be used to create an undistorted image using the approach described above.
Advantages of this approach include estimation of $\delta y$ (Eq.~(\ref{eqn:dy})) from fast EPI scans, and avoiding potential inaccuracy issues in the directly acquired $B_0$ maps.  
\iftoggle{annotated}{\textcolor{blue}{However, like fieldmap based correction, TOPUP cannot easily correct for signal pile-up in regions with severe $B_0$ inhomogeneity.}\marginnote{\textcolor{blue}{RC}}}{However, like fieldmap based correction, TOPUP cannot easily correct for signal pile-up in regions with severe $B_0$ inhomogeneity.}
\iftoggle{annotated}{\textcolor{blue}{While there are some limits to how much correction is possible, this approach has the important practical advantage that it can operate on magnitude images after simultaneous multislice reconstruction and coil combination.}\marginnote{\textcolor{blue}{RC}}}{While there are some limits to how much correction is possible, this approach has the important practical advantage that it can operate on magnitude images after simultaneous multislice reconstruction and coil combination.}

\mysubsubsection{Point-spread function mapping}
Another way to obtain $\delta y(\boldsymbol{r})$ is based on measuring the point-spread function (PSF) along $y$~\cite{Robson1997MeasurementImaging,Zeng2002ImageMapping,Zaitsev2004PointCorrection}.
This can be done by performing a calibration prescan wherein one adds an additional $k_y$ encoding blip prior to the EPI readout, and repeats the measurement with different blip areas.
For a 2D EPI acquistion, this additional ``constant-time'' encoding yields a 3D dataset where the spatial profile in the constant-time encoding dimension is the PSF~\cite{Robson1997MeasurementImaging}.
The pixel shift can then be obtained as the center (peak) of the PSF~\cite{Zaitsev2004PointCorrection}.
An advantage of this approach is that the shift is estimated with the same EPI readout as the imaging sequence to be corrected, and hence experiences and accounts for the same eddy currents and concomitant gradients.

%%%%%%%%%%%%%%%%%%%%%%%%%%%%%%%%%%%%%%%%%%%%%%%%%%%%
%%%%  non-Cartesian  %%%%%%%%%%%%%%%%%%%%%%%%%
\mysubsection{Non-Cartesian}
Off-resonance effects lead to burring and other distortions for non-Cartesian acquisitions, and this is often more objectionable than the geometric distortions seen in Cartesian imaging.  Thus, it is often necessary to do some sort of off-resonance correction in order to produce clinically useful images.  This section describes common approaches for this correction that are applicable to radial line, spiral, and related k-space trajectories.
\mysubsubsection{Direct/Analytical Methods}
\myspace
\myparagraph{Conjugate phase (CP) reconstruction}
In the absence of off-resonance, the signal equation in Eq.~(\ref{eqn:signal_omega_1D_2}) can be seen as a simple Fourier transform and a good image reconstruction will be simply an inverse FFT.  For a non-Cartesian acquisition, one can implement an approximate iFFT as: 
\begin{equation}
    {\hat{m}(x,y)} = \frac{1}{M} \sum_{i} w_i  \, y_i \, e^{i 2 \pi (k_{x,i}x + k_{y,i}y)} \approx F^{-1}\{y(t)\}
        \label{eqn:noncartft}
\end{equation}
where $i$ is the index over $M$ k-space samples $y_i = y(t_i)$ at k-space locations $(k_{x,i},k_{y,i})$, and $w_i$ is the sample density compensation term that accounts for uneven distribution of k-space samples common in non-Cartesian methods.  In practice, this expression is commonly implemented by some sort of interpolation in the k-domain to a Cartesian grid (known as gridding) followed by iFFT operators or by using integrated functions like the NUFFT~\cite{Fessler2007OnMRI}. 

%%%%%%%%%%%%%%%%%%%  figure  %%%%%%%%%%%%%%%%%%%%
\begin{wrapfigure}{r}{.5\textwidth}
\centering
\includegraphics[width=.5\textwidth]{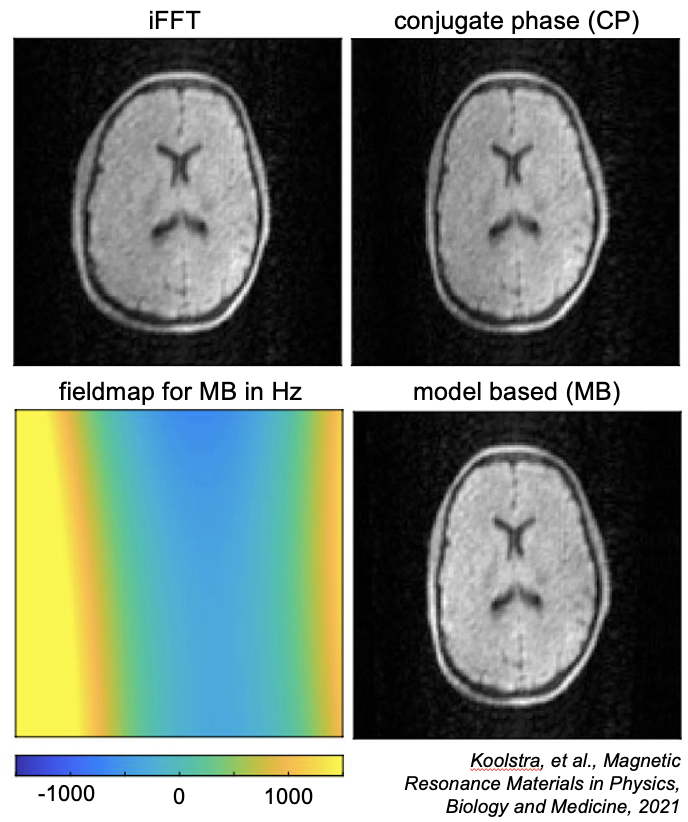}
\caption{ \footnotesize{ \textbf{CP and MB reconstruction at 50mT.} Compared to a standard iFFT reconstruction, CP and MB reconstructions both correct for geometric distortions due to off-resonance, especially in the right side of the brain where the magnetic field off-resonance is stronger. MB also shows more uniform signal intensity. \emph{(Images reused from Koolstra, et al., 2021 ~\cite{Koolstra2021ImageNonlinearities} via open access license.)}}} 
\label{fig:CPvMB}
\end{wrapfigure}

When considering off-resonance effects, a good image reconstruction may be to compensate for the off-resonance phase accumulation in the image reconstruction at every time point and for each location $(x,y)$~\cite{Macovski1985VolumetricGradients,Maeda1988ReconstructionGradients} as shown here:
\begin{equation}
    {\hat{m}(x,y)} = \frac{1}{N} \sum_{i} w_i \, y_i \, 
    \iftoggle{annotated}{\textcolor{blue}{    e^{i 2 \pi (\Delta f_0(x,y)t_i+k_{x,i}x + k_{y,i}y)} }\marginnote{\textcolor{blue}{RC}}}{    e^{i 2 \pi (\Delta f_0(x,y)t_i+k_{x,i}x + k_{y,i}y)} }
        \label{eqn:cprecon}
\end{equation}
where the
\iftoggle{annotated}{\textcolor{blue}{$\Delta f_0(x,y)t_i$ }\marginnote{\textcolor{blue}{RC}}}{$\Delta f_0(x,y)t_i$ }
term is known as the conjugate phase (CP) correction for the off-resonance (see Fig.~\ref{fig:CPvMB} for an image example). 

There are limitations to the CP reconstruction, notably in that it is not a true inverse, but rather, an approximation.

In particular, this approach fails to make high quality image reconstructions when the magnetic field 
\iftoggle{annotated}{\textcolor{blue}{($\Delta f_0(x,y)$) }\marginnote{\textcolor{blue}{RC}}}{($\Delta f_0(x,y)$)}
varies rapidly in space and therefore the ring-like PSF will also vary rapidly in space.  If two PSFs overlap in space and these PSF functions are different, then this approach cannot fully correct for both blurring functions.  An alternate way to think about this is that rapidly varying 
\iftoggle{annotated}{\textcolor{blue}{$\Delta f_0$}\marginnote{\textcolor{blue}{RC}}}{$\Delta f_0$}
corresponds to a local gradient, and therefore, the k-space is distorted for locations with large local gradients~\cite{Noll2005ConjugateCorrection}.  The distorted local k-space limits the effectiveness of the CP reconstruction.  Model-based image reconstruction, described below, is more robust to address rapid variations in the field map.

\myparagraph{Efficient approximations} 
While Eq.~(\ref{eqn:noncartft}) has fast implementations with gridding or the NUFFT,  Eq.~(\ref{eqn:cprecon}) is no longer an inverse Fourier transform due to the 
\iftoggle{annotated}{\textcolor{blue}{$\Delta f_0(x,y)t$}\marginnote{\textcolor{blue}{RC}}}{$\Delta f_0(x,y)t$}
term.  Still, there are computationally efficient implementations that result from decomposing this term into a sum of terms that vary only in space and terms that vary only in time.  The exponential can be modeled as a summation of basis functions~\cite{Fessler2005,Fessler2010h}:
\begin{equation}
    \iftoggle{annotated}{\textcolor{blue}{e^{i 2 \pi \Delta f_0(x,y) t} \approx \sum_{l=1}^{L} b_{l}(t) c_{l}(x,y)}\marginnote{\textcolor{blue}{RC}}}{e^{i 2 \pi \Delta f_0(x,y) t} \approx \sum_{l=1}^{L} b_{l}(t) c_{l}(x,y)}
    \label{eqn:b0_basis}
\end{equation}
where $b_{l}$ and $c_{l}$ represent the basis functions for the time and image space, respectively. Note that frequency, \iftoggle{annotated}{\textcolor{blue}{$\Delta f_0(x,y)$}\marginnote{\textcolor{blue}{RC}}}{$\Delta f_0(x,y)$}
, varies only with space.

The first such implementation was the so-called time-segmented approximation~\cite{Noll1991} to the CP correction term, where $b_l = b_l(t-\Bar{t}_l)$ is a family of time-varying weighting functions centered at $\bar{t}_l$ and 
\iftoggle{annotated}{\textcolor{blue}{$c_l = e^{i 2 \pi \Delta f_0(x,y) \, \bar{t}_l}$}\marginnote{\textcolor{blue}{RC}}}{$c_l = e^{i 2 \pi \Delta f_0(x,y) \, \bar{t}_l}$} is a spatially varying phase function corresponding to $\bar{t}_l$. This approximation to Eq.~(\ref{eqn:cprecon}) is:
\begin{equation}
    {\hat{m}(x,y) \approx  \sum_{l=1}^{L} F^{-1} \{ b_l(t-\bar{t}_l) \, y(t) \} 
    \iftoggle{annotated}{\textcolor{blue}{e^{i 2 \pi \Delta f_0(x,y) \, \bar{t}_l}}\marginnote{\textcolor{blue}{RC}}}{e^{i 2 \pi \Delta f_0(x,y) \, \bar{t}_l}}
    }
    \label{eqn:tscprecon}
\end{equation}
where $F^{-1}$ is the gridding or NUFFT inverse transform of windowed segments of $y(t)$.  A related form is known as the multifrequency interpolation (MFI) method~\cite{Noll1991ReconstructionImaging,Man1997MultifrequencyCorrection}, which modulates the acquired data at frequencies $f_l$ using  $b_l = e^{i 2 \pi f_l \, t}$ and interpolates between modulated images using $c_{l}(x,y)$.  This approximation to the CP reconstruction is: 
\begin{equation}
    {\hat{m}(x,y) \approx  \sum_{l=1}^{L} F^{-1} \{ e^{-i 2 \pi f_l \, t} \, y(t) \} \, c_l(
    \iftoggle{annotated}{\textcolor{blue}{\Delta f_0(x,y)}\marginnote{\textcolor{blue}{RC}}}{\Delta f_0(x,y)}
    -f_l)}.
    \label{eqn:mficprecon}
\end{equation}
Observe that $L$ Fourier transforms are implemented in both of these approximations, so while they are about $L$ times slower than Eq.~(\ref{eqn:noncartft}), they are still substantially faster than the discrete Fourier transform implementation of Eq.~(\ref{eqn:cprecon}).

Both the time-segmented approximation and the MFI methods perform similarly and can be viewed as low-rank approximations to the 
\iftoggle{annotated}{\textcolor{blue}{$e^{i 2 \pi \Delta f_0(x,y)t}$}\marginnote{\textcolor{blue}{RC}}}{$e^{i 2 \pi \Delta f_0(x,y)t}$}
term.  The interpolation functions can be selected optimally according to different optimality criteria ~\cite{Man1997MultifrequencyCorrection,Fessler2005}.  

\myparagraph{Deblurring with deconvolution}
While the above approaches implement off-resonance correction as part of the image reconstruction, this can also be implemented in the image domain using deconvolution~\cite{Ahunbay2000RapidImages}.  
Provided the spatial extent of the PSF or blurring function is small, one can deblur the images using a space variant deconvolution, where the deconvolution kernel is guided by the acquired fieldmap.  In Ahunbay et al.~\cite{Ahunbay2000RapidImages}, it was shown that in some circumstances the deconvolution kernel is separable in $x$ and $y$, which can greatly reduce the computational burden.

\myparagraph{Simulated phase evolution rewinding (SPHERE)}
An alternate approach to field correction that is similar to the CP approach is to apply the CP correction to the forward Fourier transform rather than the inverse transform.  This general approach is known as simulated phase evolution rewinding (SPHERE)~\cite{Kadah1997SimulatedImages}.  The approach starts with a reconstruction (iFFT) that does not apply an off-resonance phase correction, e.g., Eq.~(\ref{eqn:noncartft}), and follows it with a forward transform back to the k-domain, however, during the forward Fourier transform, the conjugate phase correction is applied:
\begin{equation}
  {y_{corr}(t) = \sum_{x,y} {\hat{m}(x,y)}  \, 
  \iftoggle{annotated}{\textcolor{blue}{e^{-i 2 \pi (\Delta f_0(x,y)t+[k_{x}(t)x + k_{y}(t)y])}}\marginnote{\textcolor{blue}{RC}}}{e^{-i 2 \pi (\Delta f_0(x,y)t+[k_{x}(t)x + k_{y}(t)y])}}
   }
  \label{eqn:sphere}
\end{equation}
where the sign of the complex exponent 
\iftoggle{annotated}{\textcolor{blue}{$-\Delta f_0(x,y)$ }\marginnote{\textcolor{blue}{RC}}}{$-\Delta f_0(x,y)$ }
is opposite of that in the signal equation Eq.~(\ref{eqn:signal_omega_1D_2}), thus applying the conjugate phase to compensate for the off-resonance phase accumulation.  This is followed by a simple iFFT to produce the final image.  As with the CP reconstruction, the 
\iftoggle{annotated}{\textcolor{blue}{\marginnote{\textcolor{blue}{RC}}$\Delta f_0(x,y)t$}}{$\Delta f_0(x,y)t$}
term means the forward transform is no longer a Fourier transform, but the time segmented approximations and the MFI can be used to implement this in a computationally efficient manner. Schomberg's review~\cite{Schomberg1999Off-resonanceImages} provides a detailed mathematical analysis of CP and SPHERE methods, along with fast implementation strategies and timing comparisons of the different approaches. 

\mysubsubsection{Autofocusing}
One unique aspect of non-Cartesian imaging is that the PSF may allow for autodetection of off-resonance based on features of the image.  Conceptually, it is like an autofocus camera where different parts of the image have different off-resonance corrections to produce a fully deblurred image.  This is done by examining image features rather than through the use of a fieldmap.  In~\cite{Noll1992}, a multi-frequency reconstruction was used, in which non-Cartesian images were reconstructed to produce a family of images where different parts of each image are unblurred. An image metric based on a measure of how closely the phase of low spatial frequency images aligned with the phase of high spatial frequencies was used to determine, for each voxel $(x,y)$, which modulated image was sharpest.  This measure works well for spiral imaging because the low spatial frequencies are acquired earlier than high spatial frequencies, and off-resonance will lead to these components being out of phase.  In recent work~\cite{Lim2019DynamicSpeech}, real time speech MRI with spiral readouts were automatically deblurred using a sharpness metric of features in the vocal tract. Interestingly, these autofocus methods can produce a fieldmap based on the selected image.

\myparagraph{Machine Learning} 
One challenge to autofocusing is the selection of an appropriate measure of image sharpness.  Because of the predictable and reproducible nature of the off-resonance effects in non-Cartesian imaging, it is possible to train a neural network to learn and remove image blurs.  In~\cite{Zeng2019}, body imaging with 3D cones trajectories and long readouts were deblurred by a convolution neural network (CNN) that directly maps blurred images to unblurred images acquired with shorter readouts (but an overall longer scan time).  This type of approach has also been applied to speech MRI, where a longer readout image was deblurred using a CNN, allowing for longer k-space readouts and therefore higher temporal visualization of speech in the hard to correct areas around the mouth and sinuses~\cite{Lim2020DeblurringNetworksb}. In another approach, a CNN is trained to estimate the fieldmap directly, and then uses the fieldmap within a model based image reconstruction to produce unblurred images~\cite{Haskell2022FieldMapNetReconstruction}.

%%%%%%%%%%%%%%%%%%%%%%%%%%%%%%%%%%%%%%%%%%%%%%%%%%%%%%
%%%%%%%%% model based
\mysubsection{Model based image reconstruction incorporating \texorpdfstring{\emph{B}\textsubscript{0}}{} }\label{subsec:mbir}
As the use of techniques such as parallel imaging and compressed sensing has grown, performing MRI reconstruction using model based image reconstruction (MBIR)~\cite{Fessler2020OptimizationAlgorithms} has grown in popularity. MBIR has the advantage that it can incorporate additional operations into a model of the MR imaging system beyond standard Fourier encoding, and many groups have found that incorporating a fieldmap into their imaging model can greatly aid in the reduction of $B_0$ artifacts. MBIR can also serve as the foundation for hybrid physics+machine learning approaches that combine an imaging physics model with learned parameters, often referred to as ``unrolled" algorithms~\cite{Liang2020DeepNetworks}. 

In MBIR, a so called ``forward model" is created that mathematically describes the transform from an image in vector form, denoted as $\boldsymbol{x}$, to acquired data, $\boldsymbol{y}$, using a matrix that here we will denote as $\boldsymbol{A}$. This can be thought of as a linear algebra representation of Eq.~(\ref{eqn:signal}) where data from all timepoints has been combined into a single vector $\boldsymbol{y}$, and can be written as:
\begin{equation}
    \boldsymbol{y} = \boldsymbol{A}\boldsymbol{x}
\end{equation}
Note that we have changed the letters representing the image and data from Eq.~(\ref{eqn:signal}) to align with common linear algebra nomenclature. In the simplest optimization approach, once we have a forward model, we can then optimize for the image using a linear least squares fit (since the dominant noise in MRI is Gaussian) as follows:
\begin{equation}
    \boldsymbol{\hat{x}} = \mathop{\mathrm{arg\,min}}_{\boldsymbol{x}} \frac{1}{2} \| \boldsymbol{A} \boldsymbol{x}-\boldsymbol{y} \|_2^2+\lambda R(\boldsymbol{x})  
    \label{eqn:MBIR}
\end{equation}
where $R(\boldsymbol{x})$ is a regularizer chosen by the user, and $\lambda$ is a tuning parameter to balance between the data fit term and the regularization. A common choice of regularizer in MRI is a total variation (TV) term, but there are many different options for regularizers and alternative optimization algorithms to linear least squares, as discussed in a recent overview by Fessler~\cite{Fessler2020OptimizationAlgorithms}.

In MRI, the simplest possible forward model would be to have $\boldsymbol{A}$ equal a discrete Fourier transform matrix. Additional common terms that are added include multicoil sensitivity maps and undersampling operators. The addition of an off-resonance field map is more complex, as the way the off-resonance affects the encoding forward model changes with time, as is shown in the first exponential term in Eq.~(\ref{eqn:signal_omega}), and it requires approximating the full forward model and applying corresponding optimization algorithms to leverage the efficiencies of these models. \iftoggle{annotated}{\textcolor{blue}{Solving a minimization problem of the form shown in this section, as well as the direct methods discussed in Section 5.3.1, are often referred to as ``inverse problems", and there are extensive prior work and tools available for these types of problems that can be applied to MRI.}\marginnote{\textcolor{blue}{RC}}}{Solving a minimization problem of the form shown in this section, as well as the direct methods discussed in Section 5.3.1, are often referred to as ``inverse problems", and there are extensive prior work and tools available for these types of problems that can be applied to MRI.}
\mysubsubsection{Efficient approximation of an expanded \texorpdfstring{\emph{B}\textsubscript{0}}{} forward model}
To fully model the off-resonance exponential term in Eq.~(\ref{eqn:signal_omega}) in matrix form would be computationally prohibitive, but lower order approximations of the exponential can be made to efficiently model off-resonance~\cite{Sutton2003}. As described in prior work~\cite{Fessler2005,Fessler2010h} and Eq.~(\ref{eqn:b0_basis}), the exponential can be modeled as a summation of basis functions.  Using discretized versions of the $L$ basis functions ($b_{il}$ and $c_{lj}$), the matrix vector multiplication of $\boldsymbol{Ax}$ at row $i$ (i.e., k-space data point $i$) now becomes:
\begin{equation}
    [\boldsymbol{A(\boldsymbol{\omega})x}]_i \approx \sum_{l=1}^{L} b_{il} \left[\sum_{j=1}^{N} (x_j c_{lj})e^{-i2\pi (\vec{k}_i \cdot \vec{r}_j)} \right]
    \label{eqn:A_approx}
\end{equation}
where $x_j$ is the magnetization (image value) at voxel index $j$, $N$ is the number of voxels, $\vec{k}_i \cdot \vec{r}_j$ is the inner product of the $i^{\text{th}}$ k-space sampling location with the image space location of the $j^{\text{th}}$ voxel index, and $\boldsymbol{A}(\boldsymbol{\omega})$ denotes that $\boldsymbol{A}$ now contains the fieldmap. Here, $\boldsymbol{A}(\boldsymbol{\omega})$ is implemented efficiently with $L$ NUFFTs. Additionally one can expand this to jointly optimize for the image and the fieldmap, assuming there is enough temporal variation in the k-space samples:
\begin{equation}
    \boldsymbol{\hat{x},\hat{\omega}} = \mathop{\mathrm{arg\,min}}_{\boldsymbol{x,\omega}} \frac{1}{2} \| \boldsymbol{A}(\boldsymbol{\omega}) \boldsymbol{x}-\boldsymbol{y} \|_2^2+\lambda_1 R(\boldsymbol{x})+\lambda_2 R(\boldsymbol{\omega})
    \label{eqn:MBIR_b0}
\end{equation}
Sutton et al. performed this type of joint optimization using a spiral-in/spiral-out k-space trajectory~\cite{Sutton2004} to calculate the fieldmap at every timepoint in an fMRI timeseries, allowing for dynamic $B_0$ correction of respiration induced phase oscillations. Recent work has also incorporated multicoil information to aid in the joint reconstruction of the fieldmap and image using undersampled spiral-out and EPI trajectories~\cite{Patzig2021Off-ResonanceB0-Encoding}.

Other works have expanded the forward model $\boldsymbol{A}(\boldsymbol{\omega})$ to allow for a more accurate imaging model. Joint optimization of the fieldmap and $R_2^*$ values has been performed to acquire quantitative BOLD information~\cite{Olafsson2008}. The imaging model can also be expanded to describe the differences in signal between water and fat, and to jointly optimize for water images, fat images, and the $B_0$ map~\cite{Hernando2008JointMap}. More recently, a method to model through-voxel dephasing has been presented, that uses the gradient of the fieldmap within each voxel to estimate dephasing and recover lost signal~\cite{Lam2020IntravoxelOperator}.

In addition to joint optimization, it is also possible to incorporate information from external sensors such as NMR probes to dynamically model the $B_0$ changes that occur during an MRI scan~\cite{Wilm2011HigherPerturbations}. This dynamic $B_0$ tracking approach has recently allowed the acquisition of 0.8mm $\times$ 0.8mm resolution brain imaging with a 23cm FOV at 7T by using field probes during the 57ms long single-shot spiral readouts~\cite{Kasper2022AdvancesAcquisition}, which would produce serve artifacts without $B_0$ modeling. 

Even with fast approximations, it should be noted that MBIR generally takes longer to run than direct approaches such as CP due to its iterative nature, and can also be more complex to implement. 
\iftoggle{annotated}{\textcolor{blue}{That said, the potential improvement can be substantial, and the utility of MBIR's improved accuracy will depend on the application and requirements of the imaging protocol (i.e., reconstruction time).}\marginnote{\textcolor{blue}{RC}}}{That said, the potential improvement can be substantial, and the utility of MBIR's improved accuracy will depend on the application and requirements of the imaging protocol (i.e., reconstruction time).}

\mysubsubsection{Model based reconstruction for low field}

Model based image reconstructions have also proven valuable at low field, including recently published results showing improved \emph{in vivo} human brain images from a portable low field scanners using MBIR~\cite{Cooley2021ABrain,Koolstra2021ImageNonlinearities}. In Koolstra et al., the forward model includes a $B_0$ fieldmap in a conjugate phase reconstruction (CP) as well as in a model based (MB) reconstruction (see Fig.~\ref{fig:CPvMB}). Additionally, they model gradient nonlinearities. While not the focus of this paper, gradient nonlinearities can be important to incorporate at low field, and those are actually the terms that are modeled in the MBIR of the low field work by Cooley et al., 2021~\cite{Cooley2021ABrain} not the $B_0$ field itself. Inclusion of gradient nonlinearities into the model will affect the k-space encoding term in equation \ref{eqn:A_approx}, and shows the flexibility of the MBIR approach and the benefits of using it at low field.

%%%%%%%%%%%%%%%%%%%%%%%%%%%%%%%%%%%%%%%%%%%%%
%%%%%%%%%  dynamic
\mysubsection{Dynamic correction and motion considerations} \label{subsec:dynam}

Dynamic off-resonance correction models the time-varying changes in the $B_0$ field during a scan, modeling the temporal updates either using a low rank spatial representation of the field (i.e., spherical harmonics) or an entire fieldmap. Dynamic $B_0$ correction can be done either prospectively, where the $B_0$ field is updated during the scan (this is often referred to as dynamic shimming) or retrospectively, where different frames (e.g., individual frames in an fMRI timeseries or diffusion images at different $b$-values) are reconstructed assuming a different fieldmap. 

For prospective correction, dynamic shimming can be done either in a volumetric approach, where the shim is updated during the scan to optimize homogeneity over the whole region of interest (this will require real time measurement of the $B_0$ field changes), or in a slice-by-slice manner where the optimal shims for each slice are determined, and then the shims are updated during the scan depending on which slice is being acquired (one can measure and dynamically update slice shims in real time, or optimal slice shims are determined prior to the experiment). Prospective correction without spatial information can also be useful to dynamically measure and update the center $B_0$ frequency in real time, for example in applications such as interventional MRI~\cite{Campbell-Washburn2016Real-timeFunction} and Oscillating Steady State Imaging (OSSI) fMRI~\cite{Cao2020Real-TimeFMRI}, which has been shown to be very $B_0$-sensitive~\cite{Guo2020OscillatingMRI}.

In terms of hardware to update the shims, one option is to use built in scanner shim coils (separate from the gradient coils), which generally shim up to 1$^{\text{st}}$ or 2$^{\text{nd}}$ order but can go up to 3$^{\text{rd}}$ order on some ultra high fields systems. When using built in shim coils, eddy currents should be taken into consideration as those can cause artifacts, and there are also potentially limits on the speed at which the shims can be updated using those coils. There are also more recent methods that dynamically shim using multicoil receive arrays that can also generate a DC current~\cite{Stockmann2016AImaging,Darnell2017IntegratedShimming}, one of which by Stockmann et al. is aptly named an ``AC/DC" coil. These methods take advantage of the higher order shim fields generated from a multicoil array, due to it being located much closer to the object than scanner shim coils and having more elements to fine tune a desired field shape.

Retrospective approaches have also proven useful, and are helpful when dealing with motion during a scan. When a subject moves within the scanner, it will often result in a change of the $B_0$ fieldmap, requiring dynamic distortion correction. In diffusion MRI, techniques have been developed to correct for both motion changes and distortion changes during data acquisition~\cite{Andersson2018Susceptibility-inducedData,Hutter2018Slice-levelCorrection}. Dynamic distortion correction is also important in fMRI experiments, and many groups have shown the benefits of estimating $B_0$ dynamically, either using raw k-space data~\cite{Sutton2004}, phase measurements from the images themselves~\cite{Hahn2009ImprovingTOAST,Dymerska2018AT} or from multicoil FID measurements~\cite{Wallace2021DynamicNavigators}, and then using the information in the reconstruction. Additionally, the potential clinical benefit of dynamic off-resonance correction has been show in the application of presurgical fMRI at 7T~\cite{LimaCardoso2018TheT}.

Dynamic $B_0$ correction is also relevant in low field imaging, especially in permanent magnet designs where there can be field drift during a scan~\cite{Cooley2014,Nakagomi2019DevelopmentMagnet,OReilly2021InArray}. In these approaches, a field probe or other methods are used to monitor drift during the acquisition, and then these measurements were used during the image reconstruction. On the other end of the field spectrum, dynamic $B_0$ correction has proven useful at ultra-high field, where there are large susceptibility artifacts~\cite{Juchem2010DynamicT,Sengupta2011DynamicT,Stockmann2018InT}. 

\begin{wrapfigure}{R}{.55\textwidth}
\centering
\includegraphics[width=.55\textwidth]{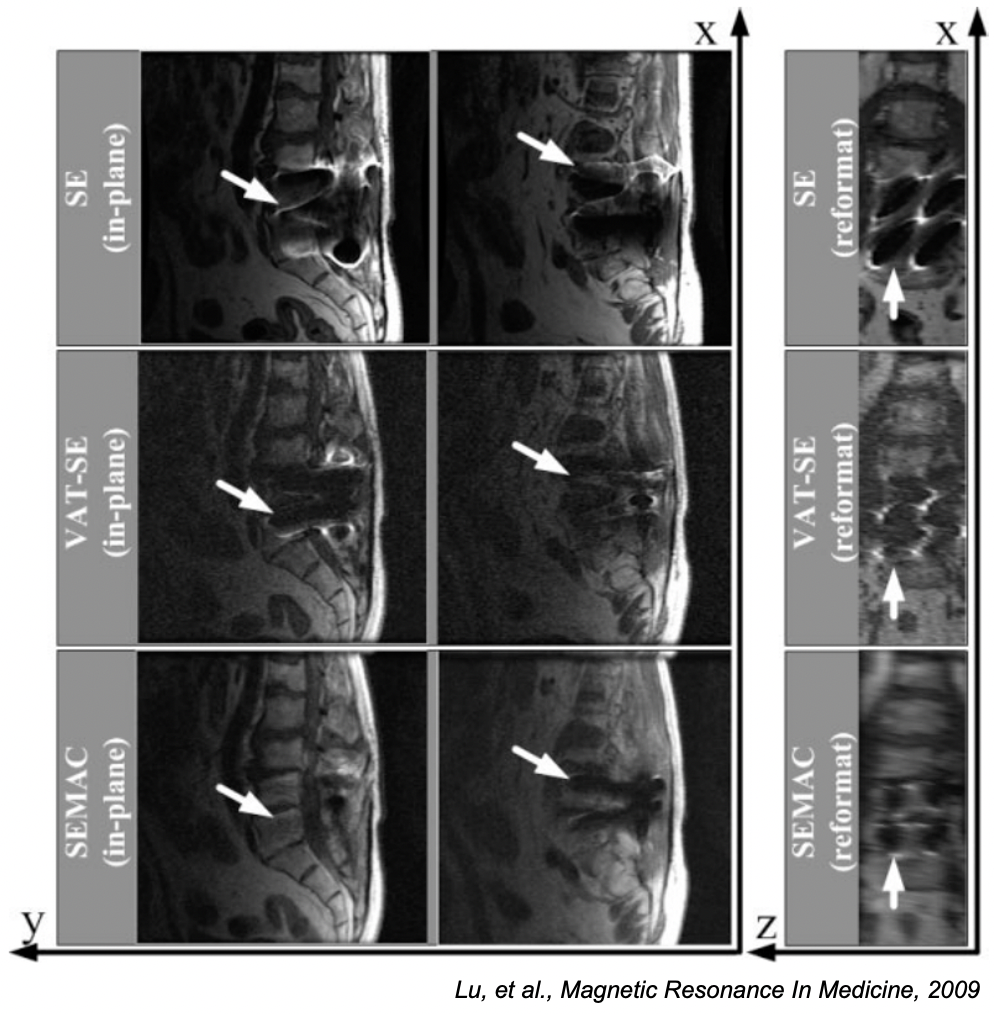}
\caption{ \footnotesize{
Comparison  of standard spin echo imaging (top row), VAT (middle), and SEMAC (bottom row) methods for a patient with a spinal implant.  The right column is reformatted so slice direction is along the horizontal axis.  The SE image is badly distorted in both frequency and slice directions, the VAT images are partially corrected in the frequency direction with substantial slice distortion, while the SEMAC image is corrected in both directions. \emph{(Images reused from~\cite{Lu2009SEMAC:MRI} with permission.)}}} 
\label{fig:semac}
\end{wrapfigure}

In MR spectroscopy, which is very sensitive to off-resonance artifacts~\cite{Juchem2021BRecommendations}, dynamic shimming is also very helpful. Spinal cord fMRI, including combined brain+spinal cord fMRI, can also benefit from dynamic shimming, because the effects of respiratory induced magnetic susceptibility changes are much more severe when imaging closer to the lungs~\cite{Finsterbusch2013CombinedUpdate,Islam2019DynamicFMRI}. Finally, in applications such as speech MRI, where the data is inherently time varying, dynamic $B_0$ correction is important, since the $B_0$ field pattern will change as a subject speaks during the exam~\cite{Lingala2016RecommendationsMRI,Lim2019DynamicSpeech,Lim2020DeblurringNetworksb}.

%%%%%%%%%%%%%%%%%%%%%%%%%%%%%%%%%%%%%%%%
%%%%%%%%%  slice shift corrections
\mysubsection{Correction for metal artifacts} \label{subsec:slice_shift_corr}

As described above, the source of $B_0$ distortions around metal implants is largely the same as other sources of susceptibility-induced $B_0$ shifts, except larger in magnitude.  Most imaging around implants uses SE (and variants) and uses traditional spin-warp type acquisitions.  This has the benefit of there being no geometric distortion in the phase encoding direction, but this still leaves artifacts in the readout and slice selection directions.  The first step to minimize artifacts is to maximize the bandwidth per resolution element in each case, by increasing the receiver bandwidth for frequency encoding readouts and increasing the RF pulse bandwidth.  While both have costs, these are useful strategies for reducing distortions. Similarly, one can use geometric distortion correction methods, like fieldmap based methods and others described above in Section \ref{subsubsec:fm_corr}, but these are ineffective for many of the larger distortions seen with metal implants.

\mysubsubsection{View angle tilting }
One clever approach for reducing in-plane distortions in the frequency direction is called view-angle tilting (VAT)~\cite{Kolind2004QuantitativeTechniques}.  Note that slices are shifted during the slice excitation, and shifted in the slice direction by an amount proportional to $B_0$-induced frequency shift.  Similarly, during frequency encoding the voxels are shifted in the readout direction proportional to that frequency shift.  The view angle tilting works by applying the slice select gradient again during frequency encoding.  If the slice gradient is the same amplitude as that used during excitation, then spins will be centered at the slice resonant frequency and not the original frequency shift.  As a result, the frequency encoding sees the correct frequency associated with each spatial location along the read direction and distortions are removed.  While effective, this approach is somewhat limited in that the frequency encoding should be about the same length as the RF pulse, which unfortunately makes the readouts very short at the expense of SNR. One could use a longer (narrow bandwidth) RF pulse, but this leads to the slices being shifted by even larger amounts, so generally, high bandwidth RF pulses and frequency encoding are used.  The main limitation of VAT is the residual and uncorrected distortion in the slice direction.

\mysubsubsection{Robust methods for distortion correction }
The extreme challenge of large $B_0$ shifts around metal implants has lead to several new methods for encoding images and the frequency offsets at the same time. One popular method is called multiacquisition variable-resonance image
combination (MAVRIC)~\cite{Koch2009AImplants}, in which slices are not excited at all, but rather narrow frequency bands are excited.  Since there is no slice selection, 3D encoding is required to prevent aliasing in slice and phase encoding directions.  This is usually combined with fast spin echo/turbo spin echo methods to improve acquisition speed. Because a narrow band of frequency is excited, there is limited distortion of the images, governed by the readout bandwidth per pixel.  To fill in the image holes, many (often overlapping) frequency bands must be excited and for each the exact shift correction for the frequency offset is applied. This creates many 3D volumes, which are then combined to create a volume that is undistorted in all three dimensions.  The disadvantage of this approach is that it has poor SNR efficiency because shifted frequency bands do not improve the SNR of other frequency bands.

As mentioned above, the main issue with VAT is that the slice distortions are not corrected.  Another comprehensive approach for distortion correction that addresses this issue is called slice-encoding for metal artifact correction
(SEMAC)~\cite{Lu2009SEMAC:MRI}, in which 2D slices are excited, but for each a 3D acquisitions is acquired so the distorted 3D profile of each excited slice can be visualized.  VAT is used to correct for in-plane distortion and the 3D shifts are spliced together into a fully corrected 3D volume.  As with MAVRIC, the SNR efficiency of this method is also poor. Figure \ref{fig:semac} demonstrates how regular spin-echo images are badly distorted in both in-plane and slice directions, how VAT mostly corrects for the in-plane distortions but not the slice direction  effects, and that SEMAC corrects for both in-plane and slice distortions.  There are MAVRIC-SEMAC hybrid approaches~\cite{Koch2011ImagingHybrid.} as well that can improve image quality and efficiency.

\mysection{Conclusions}\label{sec:conclusion}

Off-resonance artifacts are widespread in MRI, but fortunately there are many strategies to reduce and correct for them. Here we have described how off-resonance can come from a variety of sources, such as inherent $B_0$ magnet inhomogeneity, magnetic susceptibility of tissue, chemical shift, and metallic implants. The effects of these inhomogeneities depend on the type of sequence used (spin echo vs. gradient echo, Cartesian vs. non-Cartesian) and therefore the correction strategies must be different as well. We outlined the most common strategies for $B_0$ correction in Cartesian sequences (e.g., EPI), including fieldmap-based, TOPUP, and PSF approaches, as well as common strategies for non-Cartesian sequences (e.g., spiral), including direct (e.g., conjugate phase) and autofocusing methods. The effects and correction strategies for metallic implants also require a special approach, because of the very large magnetic susceptibility difference between metal and surrounding tissue. We also discussed how model based image reconstruction (MBIR) and dynamic correction can be used to correct for off-resonance in multiple types of sequences and applications. Throughout we highlighted the ways that low field MRI can be easier to work with in terms of off-resonance effects (less magnetic susceptibility and chemical shift effects), how it can be more challenging (nonuniform $B_0$ field), and the substantial progress that has been made in correcting for $B_0$ effects at low field. Overall, the type of off-resonance artifact reduction and/or correction strategy will depend on the application, and future developments in the field of off-resonance correction will continue to enable new applications and use cases for MRI.

%%%%%%%%%%   end matter      %%%%%%%%%% 
\vspace{-12pt}
\section*{Acknowledgements}
\begin{spacing}{1}
\vspace{-7pt}
The authors would like to thank Dr. Jeffrey Fessler for helpful discussions on model based image reconstructions. 
Images from Figure \ref{fig:b0_brain} are reprinted from Neuroimage, Vol 168, Stockmann \& Wald, ``In vivo B$_0$ field shimming methods for MRI at 7 T", Pages 71-87, Copyright 2018, with permission from Elsevier. 
Images from Figure \ref{fig:topup} are reprinted from Neuroimage, Vol 20, Andersson et al., ``How to correct susceptibility distortions in spin-echo echo-planar images: application to diffusion tensor imaging", Pages 870-888, Copyright 2003, with permission from Elsevier. 
Images from Figure \ref{fig:semac} are reprinted from Magnetic Resonance in Medicine, Vol 62, Issue 1, Lu et al., ``SEMAC: Slice encoding for metal artifact correction in MRI", Pages 66-76, Copyright 2009, with permission from John Wiley and Sons. This work was supported by the National Institutes of Health Grants F32EB029289, R01EB023618, U01EB026977, and U24NS120056. 
\end{spacing}

\vspace{-12pt}
\section*{Competing Interests}
\begin{spacing}{1}
\vspace{-7pt}
Melissa W. Haskell is employed by Hyperfine, Inc.
\end{spacing}

\bibliographystyle{unsrt}
\begin{multicols}{2}
\begin{spacing}{1}
{\footnotesize\bibliography{references}}

\begin{thebibliography}{100}

\bibitem{Ludeke1985SusceptibilityImaging}
K.M. L{\"{u}}deke, P.~R{\"{o}}schmann, and R.~Tischler.
\newblock {Susceptibility artefacts in NMR imaging}.
\newblock {\em Magnetic Resonance Imaging}, 3(4):329--343, 1 1985.

\bibitem{ODonnell1985NMRNonlinearities}
M.~O'Donnell and W.~A. Edelstein.
\newblock {NMR imaging in the presence of magnetic field inhomogeneities and
  gradient field nonlinearities}.
\newblock {\em Medical Physics}, 12(1):20--26, 1 1985.

\bibitem{Czervionke1988MagneticImaging.}
L.~F. Czervionke, D.~L. Daniels, F.~W. Wehrli, L.~P. Mark, L.~E. Hendrix, J.~A.
  Strandt, A.~L. Williams, and V.~M. Haughton.
\newblock {Magnetic susceptibility artifacts in gradient-recalled echo MR
  imaging.}
\newblock {\em AJNR. American journal of neuroradiology}, 9(6):1149--55, 1988.

\bibitem{Michiels1994OnNeurosurgery}
J.~Michiels, H.~Bosmans, P.~Pelgrims, D.~Vandermeulen, J.~Gybels, G.~Marchal,
  and P.~Suetens.
\newblock {On the problem of geometric distortion in magnetic resonance images
  for stereotactic neurosurgery}.
\newblock {\em Magnetic Resonance Imaging}, 12(5):749--765, 1994.

\bibitem{Ladd1996BiopsyArtifacts}
Mark~E. Ladd, Peter Erhart, Jörg~F. Debatin, Benjamin~J. Romanowski, Peter
  Boesiger, and Graeme~C. McKinnon.
\newblock {Biopsy needle susceptibility artifacts}.
\newblock {\em Magnetic Resonance in Medicine}, 36(4):646--651, 10 1996.

\bibitem{Walker2014MRIPlanning}
Amy Walker, Gary Liney, Peter Metcalfe, and Lois Holloway.
\newblock {MRI distortion: considerations for MRI based radiotherapy treatment
  planning}.
\newblock {\em Australasian Physical {\&} Engineering Sciences in Medicine},
  37(1):103--113, 3 2014.

\bibitem{Weygand2016SpatialDistortion}
Joseph Weygand, Clifton~David Fuller, Geoffrey~S. Ibbott, Abdallah~S.R.
  Mohamed, Yao Ding, Jinzhong Yang, Ken~Pin Hwang, and Jihong Wang.
\newblock {Spatial precision in magnetic resonance imaging-guided radiation
  therapy: The role of geometric distortion}.
\newblock {\em International Journal of Radiation Oncology Biology Physics},
  95(4):1304--1316, 2016.

\bibitem{Hargreaves2011Metal-inducedMRI}
Brian~A. Hargreaves, Pauline~W. Worters, Kim~Butts Pauly, John~M. Pauly,
  Kevin~M. Koch, and Garry~E. Gold.
\newblock {Metal-induced artifacts in MRI}.
\newblock {\em American Journal of Roentgenology}, 197(3):547--555, 2011.

\bibitem{Jezzard1999SourcesData}
Peter Jezzard and Stuart Clare.
\newblock {Sources of distortion in functional MRI data}.
\newblock {\em Human Brain Mapping}, 8(2-3):80--85, 1999.

\bibitem{Jezzard2012CorrectionData}
Peter Jezzard.
\newblock {Correction of geometric distortion in fMRI data}.
\newblock {\em NeuroImage}, 62(2):648--651, 2012.

\bibitem{Hahn1950SpinEchoes}
E~L Hahn.
\newblock {Spin Echoes}.
\newblock {\em Physical Review}, 80(4):580--594, 11 1950.

\bibitem{Hennig1986}
J~Hennig, A~Nauerth, and H~Friedburg.
\newblock {RARE imaging: a fast imaging method for clinical MR.}
\newblock {\em Magnetic Resonance in Medicine}, 3(6):823--833, 1986.

\bibitem{Larkman2004}
David~J Larkman, David Atkinson, and Jo~V Hajnal.
\newblock {Artifact Reduction Using Parallel Imaging Methods}.
\newblock {\em Topics in Magnetic Resonance Imaging}, 15(4):267--275, 2004.

\bibitem{Feinberg2013Ultra-fastImaging}
David~A. Feinberg and Kawin Setsompop.
\newblock {Ultra-fast MRI of the human brain with simultaneous multi-slice
  imaging}.
\newblock {\em Journal of Magnetic Resonance}, 229:90--100, 4 2013.

\bibitem{Stehling1991Echo-PlanarSecond}
Michael~K Stehling, Robert Turner, Peter Mansfield, and M~K Stehling.
\newblock {Echo-Planar Imaging: Magnetic Resonance Imaging in a Fraction of a
  Second}.
\newblock Technical report, 1991.

\bibitem{Kwong1992DynamicStimulation}
K.~K. Kwong, J.~W. Belliveau, D.~A. Chesler, I.~E. Goldberg, R.~M. Weisskoff,
  B.~P. Poncelet, D.~N. Kennedy, B.~E. Hoppel, M.~S. Cohen, R.~Turner, H.~M.
  Cheng, T.~J. Brady, and B.~R. Rosen.
\newblock {Dynamic magnetic resonance imaging of human brain activity during
  primary sensory stimulation}.
\newblock {\em Proceedings of the National Academy of Sciences of the United
  States of America}, 89(12):5675--5679, 1992.

\bibitem{Liu2015Susceptibility-weightedBrain}
Chunlei Liu, Wei Li, Karen~A. Tong, Kristen~W. Yeom, and Samuel Kuzminski.
\newblock {Susceptibility-weighted imaging and quantitative susceptibility
  mapping in the brain}.
\newblock {\em Journal of Magnetic Resonance Imaging}, 42(1):23--41, 7 2015.

\bibitem{Duyn2017ContributionsTissue}
Jeff~H. Duyn and John Schenck.
\newblock {Contributions to magnetic susceptibility of brain tissue}.
\newblock {\em NMR in Biomedicine}, 30(4), 2017.

\bibitem{Nakagomi2019DevelopmentMagnet}
Mayu Nakagomi, Michiru Kajiwara, Jumpei Matsuzaki, Katsumasa Tanabe, Sodai
  Hoshiai, Yoshikazu Okamoto, and Yasuhiko Terada.
\newblock {Development of a small car-mounted magnetic resonance imaging system
  for human elbows using a 0.2 T permanent magnet}.
\newblock {\em Journal of Magnetic Resonance}, 304:1--6, 7 2019.

\bibitem{Liu2021AScanner}
Yilong Liu, Alex~T.L. Leong, Yujiao Zhao, Linfang Xiao, Henry~K.F. Mak,
  Anderson Chun~On Tsang, Gary~K.K. Lau, Gilberto~K.K. Leung, and Ed~X. Wu.
\newblock {A low-cost and shielding-free ultra-low-field brain MRI scanner}.
\newblock {\em Nature Communications}, 12(1):1--14, 2021.

\bibitem{Cooley2021ABrain}
Clarissa~Z. Cooley, Patrick~C. McDaniel, Jason~P. Stockmann, Sai~Abitha
  Srinivas, Stephen~F. Cauley, Monika {\'{S}}liwiak, Charlotte~R. Sappo,
  Christopher~F. Vaughn, Bastien Guerin, Matthew~S. Rosen, Michael~H. Lev, and
  Lawrence~L. Wald.
\newblock {A portable scanner for magnetic resonance imaging of the brain}.
\newblock {\em Nature Biomedical Engineering}, 5(3):229--239, 2021.

\bibitem{OReilly2021InArray}
Thomas O’Reilly, Wouter~M. Teeuwisse, Danny Gans, Kirsten Koolstra, and
  Andrew~G. Webb.
\newblock {In vivo 3D brain and extremity MRI at 50 mT using a permanent magnet
  Halbach array}.
\newblock {\em Magnetic Resonance in Medicine}, 85(1):495--505, 1 2021.

\bibitem{Sheth2021AssessmentPatients}
Kevin~N. Sheth, Mercy~H. Mazurek, Matthew~M. Yuen, Bradley~A. Cahn, Jill~T.
  Shah, Adrienne Ward, Jennifer~A. Kim, Emily~J. Gilmore, Guido~J. Falcone,
  Nils Petersen, Kevin~T. Gobeske, Firas Kaddouh, David~Y. Hwang, Joseph
  Schindler, Lauren Sansing, Charles Matouk, Jonathan Rothberg, Gordon Sze,
  Jonathan Siner, Matthew~S. Rosen, Serena Spudich, and W.~Taylor Kimberly.
\newblock {Assessment of Brain Injury Using Portable, Low-Field Magnetic
  Resonance Imaging at the Bedside of Critically Ill Patients}.
\newblock {\em JAMA Neurology}, 78(1):41--47, 2021.

\bibitem{Geethanath2019AccessibleReview}
Sairam Geethanath and John~Thomas Vaughan.
\newblock {Accessible magnetic resonance imaging: A review}.
\newblock {\em Journal of Magnetic Resonance Imaging}, 49(7):e65--e77, 6 2019.

\bibitem{Fessler2020OptimizationAlgorithms}
Jeffrey~A. Fessler.
\newblock {Optimization Methods for Magnetic Resonance Image Reconstruction:
  Key Models and Optimization Algorithms}.
\newblock {\em IEEE Signal Processing Magazine}, 37(1):33--40, 2020.

\bibitem{Sutton2003}
B.P. Sutton, D.C. Noll, and J.A. Fessler.
\newblock {Fast, iterative image reconstruction for MRI in the presence of
  field inhomogeneities}.
\newblock {\em IEEE Transactions on Medical Imaging}, 22(2):178--188, 2 2003.

\bibitem{Fessler}
J.A. Fessler.
\newblock {Michigan Image Reconstruction Toolbox,
  https://web.eecs.umich.edu/{\~{}}fessler/code}, 2022.

\bibitem{Cooley2014}
Clarissa~Zimmerman Cooley, Jason~P. Stockmann, Brandon~D. Armstrong, Mathieu
  Sarracanie, Michael~H. Lev, Matthew~S. Rosen, and Lawrence~L. Wald.
\newblock {Two-dimensional imaging in a lightweight portable MRI scanner
  without gradient coils}.
\newblock {\em Magnetic Resonance in Medicine}, 883(January 2014):872--883,
  2014.

\bibitem{Cooley2018}
Clarissa~Zimmerman Cooley, Melissa~W Haskell, Stephen~F Cauley, Charlotte
  Sappo, Cristen~D Lapierre, Christopher~G Ha, Jason~P Stockmann, and
  Lawrence~L Wald.
\newblock {Design of Sparse Halbach Magnet Arrays for Portable MRI Using a
  Genetic Algorithm}.
\newblock {\em IEEE Transactions on Magnetics}, 54(1):1--12, 1 2018.

\bibitem{OReilly2019Three-dimensionalMagnet}
T.~O'Reilly, W.M. Teeuwisse, and A.G. Webb.
\newblock {Three-dimensional MRI in a homogenous 27 cm diameter bore Halbach
  array magnet}.
\newblock {\em Journal of Magnetic Resonance}, 307:106578, 10 2019.

\bibitem{McDaniel2019TheImaging}
Patrick~C. McDaniel, Clarissa~Zimmerman Cooley, Jason~P. Stockmann, and
  Lawrence~L. Wald.
\newblock {The MR Cap: A single-sided MRI system designed for potential
  point-of-care limited field-of-view brain imaging}.
\newblock {\em Magnetic Resonance in Medicine}, 82(5):1946--1960, 2019.

\bibitem{Obungoloch2018DesignHydrocephalus}
Johnes Obungoloch, Joshua~R. Harper, Steven Consevage, Igor~M. Savukov, Thomas
  Neuberger, Srinivas Tadigadapa, and Steven~J. Schiff.
\newblock {Design of a sustainable prepolarizing magnetic resonance imaging
  system for infant hydrocephalus}.
\newblock {\em Magnetic Resonance Materials in Physics, Biology and Medicine},
  31(5):665--676, 2018.

\bibitem{Schenck1996TheKinds}
John~F. Schenck.
\newblock {The role of magnetic susceptibility in magnetic resonance imaging:
  MRI magnetic compatibility of the first and second kinds}.
\newblock {\em Medical Physics}, 23(6):815--850, 6 1996.

\bibitem{Farahani1990EffectImaging}
Keyvan Farahani, Usha Sinha, Shantanu Sinha, Lee~C.L. Chiu, and Robert~B.
  Lufkin.
\newblock {Effect of field strength on susceptibility artifacts in magnetic
  resonance imaging}.
\newblock {\em Computerized Medical Imaging and Graphics}, 14(6):409--413, 11
  1990.

\bibitem{Stockmann2018InT}
Jason~P. Stockmann and Lawrence~L. Wald.
\newblock {In vivo B 0 field shimming methods for MRI at 7 T}.
\newblock {\em NeuroImage}, 168:71--87, 3 2018.

\bibitem{Elster2022Https://mriquestions.com/what-is-susceptibility.html}
Allen~D. Elster.
\newblock {https://mriquestions.com/what-is-susceptibility.html}, 2022.

\bibitem{VanSpeybroeck2021CharacterizationSystems}
C.~D.E. Van~Speybroeck, T.~O'Reilly, W.~Teeuwisse, P.~M. Arnold, and A.~G.
  Webb.
\newblock {Characterization of displacement forces and image artifacts in the
  presence of passive medical implants in low-field (<100 mT) permanent
  magnet-based MRI systems, and comparisons with clinical MRI systems}.
\newblock {\em Physica Medica}, 84(February):116--124, 2021.

\bibitem{Hood1999ChemicalRevisited}
Maureen~N. Hood, Vincent~B. Ho, James~G. Smirniotopoulos, and Jerzy Szumowski.
\newblock {Chemical Shift: The Artifact and Clinical Tool Revisited}.
\newblock {\em RadioGraphics}, 19(2):357--371, 3 1999.

\bibitem{Babcock1985EdgeEffect}
Evelyn~E. Babcock, Libby Brateman, Jeffrey~C. Weinreb, Sherye~D. Horner, and
  Ray~L. Nunnally.
\newblock {Edge Artifacts in MR Images: Chemical Shift Effect}.
\newblock {\em Journal of Computer Assisted Tomography}, 9(2):252--257, 3 1985.

\bibitem{Dwyer1985FrequencyImaging}
Andrew~J. Dwyer, Richard~H. Knop, and D.~I. Hoult.
\newblock {Frequency Shift Artifacts in MR Imaging}.
\newblock {\em Journal of Computer Assisted Tomography}, 9(1):16--18, 1 1985.

\bibitem{Smith1991ChemicalInterface.}
R.~C. Smith, R.~C. Lange, and S.~M. McCarthy.
\newblock {Chemical shift artifact: dependence on shape and orientation of the
  lipid-water interface.}
\newblock {\em Radiology}, 181(1):225--229, 10 1991.

\bibitem{ASTM2013ATSMEnvironment.}
{ASTM}.
\newblock {ATSM F2503-13 Standard practice for marking medical devices and
  other items for safety in the magnetic resonance environment.}
\newblock Technical report, American Society for Testing and Materials
  International, West Conshohocken, PA, USA, 2013.

\bibitem{IEC2014IECEnvironment.}
{IEC}.
\newblock {IEC 62570:2014 Standard practice for marking medical devices and
  other items for safety in the magnetic resonance environment.}
\newblock Technical report, International Electrotechnical Commission, Geneva,
  Switzerland, 2014.

\bibitem{Jungmann2017AdvancesMetal}
Pia~M. Jungmann, Christoph~A. Agten, Christian~W. Pfirrmann, and Reto Sutter.
\newblock {Advances in MRI around metal}.
\newblock {\em Journal of Magnetic Resonance Imaging}, 46(4):972--991, 2017.

\bibitem{Wang2005GeometricImaging}
Deming Wang and David Doddrell.
\newblock {Geometric Distortion in Structural Magnetic Resonance Imaging}.
\newblock {\em Current Medical Imaging Reviews}, 1(1):49--60, 1 2005.

\bibitem{Macovski1996NoiseMRI}
Albert Macovski.
\newblock {Noise in MRI}.
\newblock {\em Magnetic Resonance in Medicine}, 36(3):494--497, 9 1996.

\bibitem{Porter2004Multi-shotCorrection}
D~Porter and E~Mueller.
\newblock {Multi-shot diffusion-weighted EPI with readout mosaic segmentation
  and 2D navigator correction}.
\newblock In {\em Proceedings of the 12th Annual Meeting of ISMRM, Kyoto}, page
  442, 2004.

\bibitem{Holdsworth2008Readout-segmented3T}
Samantha~J. Holdsworth, Stefan Skare, Rexford~D. Newbould, Raphael Guzmann,
  Nikolas~H. Blevins, and Roland Bammer.
\newblock {Readout-segmented EPI for rapid high resolution diffusion imaging at
  3T}.
\newblock {\em European Journal of Radiology}, 65(1):36--46, 1 2008.

\bibitem{Butts1996Diffusion-weightedEchoes}
Kim Butts, Alex de~Crespigny, John~M Pauly, and Michael Moseley.
\newblock {Diffusion-weighted interleaved echo-planar imaging with a pair of
  orthogonal navigator echoes}.
\newblock {\em Magnetic Resonance in Medicine}, 35(5):763--770, 5 1996.

\bibitem{Noll1997MultishotImaging}
Douglas~C. Noll.
\newblock {Multishot rosette trajectories for spectrally selective mr imaging}.
\newblock {\em IEEE Transactions on Medical Imaging}, 16(4):372--377, 1997.

\bibitem{Weiger2013ZTEHumans}
Markus Weiger, David~O. Brunner, Benjamin~E. Dietrich, Colin~F. M{\"{u}}ller,
  and Klaas~P. Pruessmann.
\newblock {ZTE imaging in humans}.
\newblock {\em Magnetic Resonance in Medicine}, 70(2):328--332, 8 2013.

\bibitem{Gatehouse2003MagneticTissue}
P.D. Gatehouse and G.M. Bydder.
\newblock {Magnetic Resonance Imaging of Short T2 Components in Tissue}.
\newblock {\em Clinical Radiology}, 58(1):1--19, 1 2003.

\bibitem{Du2004Contrast-enhancedReconstruction}
Jiang Du, Timothy~J. Carroll, Ethan Brodsky, Aiming Lu, T.M. Grist, Charles~A.
  Mistretta, and Walter~F. Block.
\newblock {Contrast-enhanced peripheral magnetic resonance angiography using
  time-resolved vastly undersampled isotropic projection reconstruction}.
\newblock {\em Journal of Magnetic Resonance Imaging}, 20(5):894--900, 11 2004.

\bibitem{Feng2014Golden-angleMRI}
Li~Feng, Robert Grimm, Kai~Tobias Block, Hersh Chandarana, Sungheon Kim, Jian
  Xu, Leon Axel, Daniel~K. Sodickson, and Ricardo Otazo.
\newblock {Golden-angle radial sparse parallel MRI: Combination of compressed
  sensing, parallel imaging, and golden-angle radial sampling for fast and
  flexible dynamic volumetric MRI}.
\newblock {\em Magnetic Resonance in Medicine}, 72(3):707--717, 9 2014.

\bibitem{Noll1995SpiralActivation}
Douglas~C. Noll, Jonathan~D. Cohen, Craig~H. Meyer, and Walter Schneider.
\newblock {Spiral K-space MR imaging of cortical activation}.
\newblock {\em Journal of Magnetic Resonance Imaging}, 5(1):49--56, 1 1995.

\bibitem{Nayak2005SpiralImaging}
Krishna~S. Nayak, Brian~A. Hargreaves, Bob~S. Hu, Dwight~G. Nishimura, John~M.
  Pauly, and Craig~H. Meyer.
\newblock {Spiral balanced steady-state free precession cardiac imaging}.
\newblock {\em Magnetic Resonance in Medicine}, 53(6):1468--1473, 6 2005.

\bibitem{Li2020ImprovingTechnique}
Zhiqiang Li, James~G. Pipe, Melvyn~B. Ooi, Michael Kuwabara, and John~P. Karis.
\newblock {Improving the image quality of 3D FLAIR with a spiral MRI
  technique}.
\newblock {\em Magnetic Resonance in Medicine}, 83(1):170--177, 1 2020.

\bibitem{Irarrazabal1995FastImaging}
Pablo Irarrazabal and Dwight~G. Nishimura.
\newblock {Fast Three Dimensional Magnetic Resonance Imaging}.
\newblock {\em Magnetic Resonance in Medicine}, 33(5):656--662, 5 1995.

\bibitem{Hutton2002ImageEvaluation}
Chloe Hutton, Andreas Bork, Oliver Josephs, Ralf Deichmann, John Ashburner, and
  Robert Turner.
\newblock {Image distortion correction in fMRI: A quantitative evaluation}.
\newblock {\em NeuroImage}, 16(1):217--240, 2002.

\bibitem{Reber1998CorrectionMaps}
Paul~J. Reber, Eric~C. Wong, Richard~B. Buxton, and Lawrence~R. Frank.
\newblock {Correction of off resonance-related distortion in echo-planar
  imaging using EPI-based field maps}.
\newblock {\em Magnetic Resonance in Medicine}, 39(2):328--330, 1998.

\bibitem{Jenkinson2012FSL}
Mark Jenkinson, Christian~F Beckmann, Timothy~E.J. Behrens, Mark~W Woolrich,
  and Stephen~M Smith.
\newblock {FSL}.
\newblock {\em NeuroImage}, 62(2):782--790, 8 2012.

\bibitem{Jenkinson2003FastAlgorithm}
Mark Jenkinson.
\newblock {Fast, automated, N-dimensional phase-unwrapping algorithm}.
\newblock {\em Magnetic Resonance in Medicine}, 49(1):193--197, 2003.

\bibitem{Iyer2020PhysiCal:Mapping}
Siddharth~Srinivasan Iyer, Congyu Liao, Qing Li, Mary~Katherine Manhard, Avery
  Berman, Berkin Bilgic, and Kawin Setsompop.
\newblock {PhysiCal: A rapid calibration scan for B0, B1+, coil sensitivity and
  Eddy current mapping}.
\newblock {\em Proceedings of the 28th Annual Meeting of ISMRM,
  Sydney/Virtual}, page 0661, 2020.

\bibitem{Barmet2008SpatiotemporalMR}
Christoph Barmet, Nicola De~Zanche, and Klaas~P. Pruessmann.
\newblock {Spatiotemporal magnetic field monitoring for MR}.
\newblock {\em Magnetic Resonance in Medicine}, 60(1):187--197, 2008.

\bibitem{Dietrich2016AAnalysis}
Benjamin~E. Dietrich, David~O. Brunner, Bertram~J. Wilm, Christoph BarmeT,
  Simon Gross, Lars Kasper, Maximilian Haeberlin, Thomas Schmid, S.~Johanna
  Vannesjo, and Klaas~P. Pruessmann.
\newblock {A field camera for MR sequence monitoring and system analysis}.
\newblock {\em Magnetic Resonance in Medicine}, 75(4):1831--1840, 2016.

\bibitem{Gross2016DynamicResolution}
Simon Gross, Christoph Barmet, Benjamin~E. Dietrich, David~O. Brunner, Thomas
  Schmid, and Klaas~P. Pruessmann.
\newblock {Dynamic nuclear magnetic resonance field sensing with
  part-per-trillion resolution}.
\newblock {\em Nature Communications}, 7:1--6, 2016.

\bibitem{Wallace2020RapidImage}
Tess~E. Wallace, Onur Afacan, Tobias Kober, and Simon~K. Warfield.
\newblock {Rapid measurement and correction of spatiotemporal B0 field changes
  using FID navigators and a multi-channel reference image}.
\newblock {\em Magnetic Resonance in Medicine}, 83(2):575--589, 2020.

\bibitem{Jezzard1995CorrectionVariations}
Peter Jezzard and Robert~S. Balaban.
\newblock {Correction for geometric distortion in echo planar images from B0
  field variations}.
\newblock {\em Magnetic Resonance in Medicine}, 34(1):65--73, 7 1995.

\bibitem{Funai2008RegularizedMRI}
Amanda~K. Funai, Jeffrey~A. Fessler, Desmond~T.B. Yeo, Douglas~C. Noll, and
  Valur~T. Olafsson.
\newblock {Regularized field map estimation in MRI}.
\newblock {\em IEEE Transactions on Medical Imaging}, 27(10):1484--1494, 2008.

\bibitem{Lin2020EfficientMRI}
Claire~Yilin Lin and Jeffrey~A. Fessler.
\newblock {Efficient Regularized Field Map Estimation in 3D MRI}.
\newblock {\em IEEE Transactions on Computational Imaging}, 6(1):1451--1458,
  2020.

\bibitem{Schneider1991RapidShimming}
Erika Schneider and Gary Glover.
\newblock {Rapid in vivo proton shimming}.
\newblock {\em Magnetic Resonance in Medicine}, 18(2):335--347, 4 1991.

\bibitem{Juchem2021BRecommendations}
Christoph Juchem, Cristina Cudalbu, Robin~A. Graaf, Rolf Gruetter, Anke
  Henning, Hoby~P. Hetherington, and Vincent~O. Boer.
\newblock {B 0 shimming for in vivo magnetic resonance spectroscopy: Experts'
  consensus recommendations}.
\newblock {\em NMR in Biomedicine}, 34(5):1--20, 5 2021.

\bibitem{Larkman2007a}
David~J Larkman and Rita~G Nunes.
\newblock {Parallel magnetic resonance imaging.}
\newblock {\em Physics in medicine and biology}, 52(7):15--55, 4 2007.

\bibitem{Bley2010FatImaging}
Thorsten~A. Bley, Oliver Wieben, Christopher~J. Fran{\c{c}}ois, Jean~H.
  Brittain, and Scott~B. Reeder.
\newblock {Fat and water magnetic resonance imaging}.
\newblock {\em Journal of Magnetic Resonance Imaging}, 31(1):4--18, 1 2010.

\bibitem{Andersson2003HowImaging}
Jesper~L.R. Andersson, Stefan Skare, and John Ashburner.
\newblock {How to correct susceptibility distortions in spin-echo echo-planar
  images: Application to diffusion tensor imaging}.
\newblock {\em NeuroImage}, 20(2):870--888, 2003.

\bibitem{Schallmo2021AssessingData}
Michael~Paul Schallmo, Kimberly~B. Weldon, Philip~C. Burton, Scott~R. Sponheim,
  and Cheryl~A. Olman.
\newblock {Assessing methods for geometric distortion compensation in 7 T
  gradient echo functional MRI data}.
\newblock {\em Human Brain Mapping}, 42(13):4205--4223, 2021.

\bibitem{Abreu2021QuantitativeAnalyses}
Rodolfo Abreu and João~Valente Duarte.
\newblock {Quantitative Assessment of the Impact of Geometric Distortions and
  Their Correction on fMRI Data Analyses}.
\newblock {\em Frontiers in Neuroscience}, 15(March):1--17, 3 2021.

\bibitem{Robson1997MeasurementImaging}
Matthew~D. Robson, John~C. Gore, and R.~Todd Constable.
\newblock {Measurement of the point spread function in MRI using constant time
  imaging}.
\newblock {\em Magnetic Resonance in Medicine}, 38(5):733--740, 1997.

\bibitem{Zeng2002ImageMapping}
Huairen Zeng and R.~Todd Constable.
\newblock {Image distortion correction in EPI: Comparison of field mapping with
  point spread function mapping}.
\newblock {\em Magnetic Resonance in Medicine}, 48(1):137--146, 2002.

\bibitem{Zaitsev2004PointCorrection}
M.~Zaitsev, J.~Hennig, and O.~Speck.
\newblock {Point spread function mapping with parallel imaging techniques and
  high acceleration factors: Fast, robust, and flexible method for echo-planar
  imaging distortion correction}.
\newblock {\em Magnetic Resonance in Medicine}, 52(5):1156--1166, 2004.

\bibitem{Fessler2007OnMRI}
Jeffrey~A. Fessler.
\newblock {On NUFFT-based gridding for non-Cartesian MRI}.
\newblock {\em Journal of Magnetic Resonance}, 188(2):191--195, 2007.

\bibitem{Koolstra2021ImageNonlinearities}
Kirsten Koolstra, Thomas O’Reilly, Peter B{\"{o}}rnert, and Andrew Webb.
\newblock {Image distortion correction for MRI in low field permanent magnet
  systems with strong B0 inhomogeneity and gradient field nonlinearities}.
\newblock {\em Magnetic Resonance Materials in Physics, Biology and Medicine},
  34(4):631--642, 2021.

\bibitem{Macovski1985VolumetricGradients}
Albert Macovski.
\newblock {Volumetric NMR imaging with time-varying gradients}.
\newblock {\em Magnetic Resonance in Medicine}, 2(1):29--40, 2 1985.

\bibitem{Maeda1988ReconstructionGradients}
Akira Maeda, Koichi Sano, and Tetsuo Yokoyama.
\newblock {Reconstruction by weighted correlation for MRI with time-varying
  gradients}.
\newblock {\em IEEE Transactions on Medical Imaging}, 7(1):26--31, 3 1988.

\bibitem{Noll2005ConjugateCorrection}
D.C. Noll, J.A. Fessler, and B.P. Sutton.
\newblock {Conjugate phase MRI reconstruction with spatially variant sample
  density correction}.
\newblock {\em IEEE Transactions on Medical Imaging}, 24(3):325--336, 3 2005.

\bibitem{Fessler2005}
J.A. Fessler, {Sangwoo Lee}, V.T. Olafsson, H.R. Shi, and D.C. Noll.
\newblock {Toeplitz-based iterative image reconstruction for MRI with
  correction for magnetic field inhomogeneity}.
\newblock {\em IEEE Transactions on Signal Processing}, 53(9):3393--3402, 9
  2005.

\bibitem{Fessler2010h}
Jeffrey Fessler.
\newblock {Model-Based Image Reconstruction for MRI}.
\newblock {\em IEEE Signal Processing Magazine}, 27(4):81--89, 7 2010.

\bibitem{Noll1991}
D.C. Noll, C.H. Meyer, J.M. Pauly, D.G. Nishimura, and Albert Macovski.
\newblock {A homogeneity correction method for magnetic resonance imaging with
  time-varying gradients}.
\newblock {\em IEEE Transactions on Medical Imaging}, 10(4):629--637, 1991.

\bibitem{Noll1991ReconstructionImaging}
D.~C. Noll.
\newblock {\em {Reconstruction techniques for magnetic resonance imaging}}.
\newblock PhD thesis, Stanford University, 1991.

\bibitem{Man1997MultifrequencyCorrection}
Lai-Chee Man, John~M. Pauly, and Albert Macovski.
\newblock {Multifrequency interpolation for fast off-resonance correction}.
\newblock {\em Magnetic Resonance in Medicine}, 37(5):785--792, 5 1997.

\bibitem{Ahunbay2000RapidImages}
Ergun Ahunbay and James~G. Pipe.
\newblock {Rapid method for deblurring spiral MR images}.
\newblock {\em Magnetic Resonance in Medicine}, 44(3):491--494, 9 2000.

\bibitem{Kadah1997SimulatedImages}
Yasser~M Kadah and Xiaoping Hu.
\newblock {Simulated phase evolution rewinding (SPHERE): A technique for
  reducing B0 inhomogeneity effects in MR images}.
\newblock {\em Magnetic Resonance in Medicine}, 38(4):615--627, 10 1997.

\bibitem{Schomberg1999Off-resonanceImages}
Hermann Schomberg.
\newblock {Off-resonance correction of MR images}.
\newblock {\em IEEE Transactions on Medical Imaging}, 18(6):481--495, 6 1999.

\bibitem{Noll1992}
Douglas~C. Noll, John~M. Pauly, Craig~H. Meyer, Dwight~G. Nishimura, and Albert
  Macovskj.
\newblock {Deblurring for non‐2D fourier transform magnetic resonance
  imaging}.
\newblock {\em Magnetic Resonance in Medicine}, 25(2):319--333, 1992.

\bibitem{Lim2019DynamicSpeech}
Yongwan Lim, Sajan~Goud Lingala, Shrikanth~S. Narayanan, and Krishna~S. Nayak.
\newblock {Dynamic off-resonance correction for spiral real-time MRI of
  speech}.
\newblock {\em Magnetic Resonance in Medicine}, 81(1):234--246, 2019.

\bibitem{Zeng2019}
David~Y. Zeng, Jamil Shaikh, Signy Holmes, Ryan~L. Brunsing, John~M. Pauly,
  Dwight~G. Nishimura, Shreyas~S. Vasanawala, and Joseph~Y. Cheng.
\newblock {Deep residual network for off-resonance artifact correction with
  application to pediatric body MRA with 3D cones}.
\newblock {\em Magnetic Resonance in Medicine}, 82(4):1398--1411, 10 2019.

\bibitem{Lim2020DeblurringNetworksb}
Yongwan Lim, Yannick Bliesener, Shrikanth Narayanan, and Krishna~S. Nayak.
\newblock {Deblurring for spiral real‐time MRI using convolutional neural
  networks}.
\newblock {\em Magnetic Resonance in Medicine}, 84(6):3438--3452, 12 2020.

\bibitem{Haskell2022FieldMapNetReconstruction}
Melissa~W. Haskell, Anish Lahiri, Jon-Fredrik Nielsen, Jeffrey~A. Fessler, and
  Douglas~C. Noll.
\newblock {FieldMapNet MRI: Learning-based mapping from single echo time BOLD
  fMRI data to fieldmaps with model-based reconstruction}.
\newblock {\em Proceedings of the 30th Annual Meeting of ISMRM, London}, page
  0235, 2022.

\bibitem{Liang2020DeepNetworks}
Dong Liang, Jing Cheng, Ziwen Ke, and Leslie Ying.
\newblock {Deep Magnetic Resonance Image Reconstruction: Inverse Problems Meet
  Neural Networks}.
\newblock {\em IEEE Signal Processing Magazine}, 37(1):141--151, 2020.

\bibitem{Sutton2004}
Bradley~P. Sutton, Douglas~C. Noll, and Jeffrey~A. Fessler.
\newblock {Dynamic field map estimation using a spiral-in/spiral-out
  acquisition}.
\newblock {\em Magnetic Resonance in Medicine}, 51(6):1194--1204, 6 2004.

\bibitem{Patzig2021Off-ResonanceB0-Encoding}
Franz Patzig, Bertram Wilm, and Klaas~Paul Pruessmann.
\newblock {Off-Resonance Self-Correction by Implicit B0-Encoding}.
\newblock {\em Proceedings of the 29th Annual Meeting of ISMRM, Virtual}, page
  0666, 2021.

\bibitem{Olafsson2008}
V.T. Olafsson, D.C. Noll, and J.A. Fessler.
\newblock {Fast Joint Reconstruction of Dynamic R2* and Field Maps in
  Functional MRI}.
\newblock {\em IEEE Transactions on Medical Imaging}, 27(9):1177--1188, 9 2008.

\bibitem{Hernando2008JointMap}
Diego Hernando, J.~P. Haldar, B.~P. Sutton, J.~Ma, P.~Kellman, and Z.-P. Liang.
\newblock {Joint estimation of water/fat images and field inhomogeneity map}.
\newblock {\em Magnetic Resonance in Medicine}, 59(3):571--580, 3 2008.

\bibitem{Lam2020IntravoxelOperator}
Fan Lam and Bradley~P. Sutton.
\newblock {Intravoxel B0 inhomogeneity corrected reconstruction using a
  low-rank encoding operator}.
\newblock {\em Magnetic Resonance in Medicine}, 84(2):885--894, 2020.

\bibitem{Wilm2011HigherPerturbations}
Bertram~J. Wilm, Christoph Barmet, Matteo Pavan, and Klaas~P. Pruessmann.
\newblock {Higher order reconstruction for MRI in the presence of
  spatiotemporal field perturbations}.
\newblock {\em Magnetic Resonance in Medicine}, 65(6):1690--1701, 2011.

\bibitem{Kasper2022AdvancesAcquisition}
Lars Kasper, Maria Engel, Jakob Heinzle, Matthias Mueller-Schrader, Nadine~N.
  Graedel, Jonas Reber, Thomas Schmid, Christoph Barmet, Bertram~J. Wilm,
  Klaas~Enno Stephan, and Klaas~P. Pruessmann.
\newblock {Advances in spiral fMRI: A high-resolution study with single-shot
  acquisition}.
\newblock {\em NeuroImage}, 246(November 2021):118738, 2022.

\bibitem{Campbell-Washburn2016Real-timeFunction}
Adrienne~E. Campbell-Washburn, Hui Xue, Robert~J. Lederman, Anthony~Z.
  Faranesh, and Michael~S. Hansen.
\newblock {Real-time distortion correction of spiral and echo planar images
  using the gradient system impulse response function}.
\newblock {\em Magnetic Resonance in Medicine}, 75(6):2278--2285, 6 2016.

\bibitem{Cao2020Real-TimeFMRI}
Amos~A Cao and Douglas Noll.
\newblock {Real-Time Respiration Compensation in Oscillating Steady State
  fMRI}.
\newblock {\em Proceedings of the 28th Annual Meeting of ISMRM,
  Sydney/Virtual}, page 1223, 2020.

\bibitem{Guo2020OscillatingMRI}
Shouchang Guo and Douglas~C. Noll.
\newblock {Oscillating steady‐state imaging (OSSI): A novel method for
  functional MRI}.
\newblock {\em Magnetic Resonance in Medicine}, 84(2):698--712, 8 2020.

\bibitem{Stockmann2016AImaging}
Jason~P. Stockmann, Thomas Witzel, Boris Keil, Jonathan~R. Polimeni, Azma
  Mareyam, Cristen Lapierre, Kawin Setsompop, and Lawrence~L. Wald.
\newblock {A 32-channel combined RF and B0 shim array for 3T brain imaging}.
\newblock {\em Magnetic Resonance in Medicine}, 75(1):441--451, 2016.

\bibitem{Darnell2017IntegratedShimming}
Dean Darnell, Trong~Kha Truong, and Allen~W. Song.
\newblock {Integrated parallel reception, excitation, and shimming (iPRES) with
  multiple shim loops per radio-frequency coil element for improved B0
  shimming}.
\newblock {\em Magnetic Resonance in Medicine}, 77(5):2077--2086, 2017.

\bibitem{Andersson2018Susceptibility-inducedData}
Jesper~L.R. Andersson, Mark~S. Graham, Ivana Drobnjak, Hui Zhang, and Jon
  Campbell.
\newblock {Susceptibility-induced distortion that varies due to motion:
  Correction in diffusion MR without acquiring additional data}.
\newblock {\em NeuroImage}, 171(July 2017):277--295, 2018.

\bibitem{Hutter2018Slice-levelCorrection}
Jana Hutter, Daan~J. Christiaens, Torben Schneider, Lucilio Cordero-Grande,
  Paddy~J. Slator, Maria Deprez, Anthony~N. Price, J.~Donald Tournier, Mary
  Rutherford, and Joseph~V. Hajnal.
\newblock {Slice-level diffusion encoding for motion and distortion
  correction}.
\newblock {\em Medical Image Analysis}, 48:214--229, 2018.

\bibitem{Hahn2009ImprovingTOAST}
Andrew~D. Hahn, Andrew~S. Nencka, and Daniel~B. Rowe.
\newblock {Improving robustness and reliability of phase-sensitive fMRI
  analysis using temporal off-resonance alignment of single-echo timeseries
  (TOAST)}.
\newblock {\em NeuroImage}, 44(3):742--752, 2 2009.

\bibitem{Dymerska2018AT}
Barbara Dymerska, Benedikt~A. Poser, Markus Barth, Siegfried Trattnig, and
  Simon~D. Robinson.
\newblock {A method for the dynamic correction of B 0 -related distortions in
  single-echo EPI at 7 T}.
\newblock {\em NeuroImage}, 168:321--331, 3 2018.

\bibitem{Wallace2021DynamicNavigators}
Tess~E. Wallace, Jonathan~R. Polimeni, Jason~P. Stockmann, W.~Scott Hoge,
  Tobias Kober, Simon~K. Warfield, and Onur Afacan.
\newblock {Dynamic distortion correction for functional MRI using FID
  navigators}.
\newblock {\em Magnetic Resonance in Medicine}, 85(3):1294--1307, 3 2021.

\bibitem{LimaCardoso2018TheT}
Pedro Lima~Cardoso, Barbara Dymerska, Beáta Bachrat{\'{a}}, Florian Ph~S.
  Fischmeister, Nina Mahr, Eva Matt, Siegfried Trattnig, Roland Beisteiner, and
  Simon~Daniel Robinson.
\newblock {The clinical relevance of distortion correction in presurgical fMRI
  at 7 T}.
\newblock {\em NeuroImage}, 168(December 2016):490--498, 2018.

\bibitem{Juchem2010DynamicT}
Christoph Juchem, Terence~W. Nixon, Piotr Diduch, Douglas~L. Rothman, Piotr
  Starewicz, and Robin~A. De~Graaf.
\newblock {Dynamic shimming of the human brain at 7 T}.
\newblock {\em Concepts in Magnetic Resonance Part B: Magnetic Resonance
  Engineering}, 37B(3):116--128, 7 2010.

\bibitem{Sengupta2011DynamicT}
Saikat Sengupta, E.~Brian Welch, Yansong Zhao, David Foxall, Piotr Starewicz,
  Adam~W. Anderson, John~C. Gore, and Malcolm~J. Avison.
\newblock {Dynamic B0 shimming at 7 T}.
\newblock {\em Magnetic Resonance Imaging}, 29(4):483--496, 2011.

\bibitem{Lu2009SEMAC:MRI}
Wenmiao Lu, Kim~Butts Pauly, Garry~E. Gold, John~M. Pauly, and Brian~A.
  Hargreaves.
\newblock {SEMAC: Slice encoding for metal artifact correction in MRI}.
\newblock {\em Magnetic Resonance in Medicine}, 62(1):66--76, 7 2009.

\bibitem{Finsterbusch2013CombinedUpdate}
Jürgen Finsterbusch, Christian Sprenger, and Christian B{\"{u}}chel.
\newblock {Combined T2*-weighted measurements of the human brain and cervical
  spinal cord with a dynamic shim update}.
\newblock {\em NeuroImage}, 79:153--161, 10 2013.

\bibitem{Islam2019DynamicFMRI}
Haisam Islam, Christine S.~W. Law, Kenneth~A. Weber, Sean~C. Mackey, and
  Gary~H. Glover.
\newblock {Dynamic per slice shimming for simultaneous brain and spinal cord
  fMRI}.
\newblock {\em Magnetic Resonance in Medicine}, 81(2):825--838, 2 2019.

\bibitem{Lingala2016RecommendationsMRI}
Sajan~Goud Lingala, Brad~P. Sutton, Marc~E. Miquel, and Krishna~S. Nayak.
\newblock {Recommendations for real-time speech MRI}.
\newblock {\em Journal of Magnetic Resonance Imaging}, 43(1):28--44, 2016.

\bibitem{Kolind2004QuantitativeTechniques}
Shannon~H. Kolind, Alex~L. MacKay, Peter~L. Munk, and Qing-San Xiang.
\newblock {Quantitative evaluation of metal artifact reduction techniques}.
\newblock {\em Journal of Magnetic Resonance Imaging}, 20(3):487--495, 9 2004.

\bibitem{Koch2009AImplants}
Kevin~M. Koch, John~E. Lorbiecki, R.~Scott Hinks, and Kevin~F. King.
\newblock {A multispectral three-dimensional acquisition technique for imaging
  near metal implants}.
\newblock {\em Magnetic Resonance in Medicine}, 61(2):381--390, 2009.

\bibitem{Koch2011ImagingHybrid.}
K.~M. Koch, A.~C. Brau, W.~Chen, G.~E. Gold, B.~A. Hargreaves, M.~Koff, G.~C.
  McKinnon, H.~G. Potter, and K.~F. King.
\newblock {Imaging near metal with a MAVRIC-SEMAC hybrid.}
\newblock {\em Magnetic resonance in medicine : official journal of the Society
  of Magnetic Resonance in Medicine / Society of Magnetic Resonance in
  Medicine}, 65(1):71--82, 2011.

\end{thebibliography}
\end{spacing}
\end{multicols}

\end{document}